\documentclass[english,reprint,aps,pra,superscriptaddress,longbibliography,footinbib]{revtex4-2}

\usepackage{graphicx,mathrsfs,enumerate,epstopdf}
\usepackage[utf8]{inputenc}
\usepackage{mathtools}
\usepackage{amsfonts,amstext}
\usepackage{amsmath,amsthm,amssymb,tensor,diagbox,multirow}
\usepackage{physics}
\usepackage[hidelinks]{hyperref}\hypersetup{pdfpagemode=UseNone,pdfstartview=}
\usepackage[left=2cm,right=2cm]{geometry}
\usepackage{xcolor}
\usepackage{xfrac}
\usepackage{url}
\usepackage{times}
\usepackage{soul}
\usepackage{dsfont}
\usepackage[ruled,vlined,linesnumbered]{algorithm2e}
\usepackage{complexity}
\usepackage{tikz}
\usepackage{makecell}
\usepackage{tabularray}
\usepackage{comment}

\makeatletter

\SetKwComment{Comment}{/*$\,$}{}

\SetCommentSty{mycommfont}

\newtheorem{dfn}{Definition}
\newtheorem{thm}{Theorem}

\newtheorem{lem}[thm]{Lemma}

\newcommand{\suc}{\text{suc}}

\newcommand{\TII}{\affiliation{Quantum Research Centre, Technology Innovation Institute, Abu Dhabi,
UAE}}

\newclass{\DQC}{DQC1}

\usepackage[caption=false]{subfig}

\makeatother

\usepackage{babel}

\begin{document}
\title{Partition function estimation with a quantum coin toss}
\author{Thais L. Silva}
\TII

\author{Lucas Borges}
\TII
\affiliation{Federal University of Rio de Janeiro, Caixa Postal 68528, Rio de Janeiro, RJ 21941-972, Brazil}
 
\author{Leandro Aolita}
\TII

\begin{abstract}
Estimating quantum partition functions is a critical task in a variety of fields.
 However, the problem is classically intractable in general due to the exponential scaling of the Hamiltonian dimension $N$ in the number of particles. This paper introduces a quantum algorithm for estimating the partition function $Z_\beta$ of a
generic Hamiltonian $H$ up to multiplicative error based on a quantum coin toss.
The coin is defined by the probability of applying the quantum imaginary-time evolution propagator $f_\beta[H]=e^{-\beta H/{2}}$ at inverse temperature $\beta$ to the maximally mixed state, 
realized by a block-encoding of $f_\beta[H]$ into a unitary quantum circuit followed by a post-selection measurement.
 Our algorithm does not use costly subroutines such as quantum phase estimation or amplitude amplification; and the binary nature of the coin allows us to invoke tools from Bernoulli-process analysis to prove a runtime scaling as $\mathcal{O}(N/{Z_\beta})$, quadratically better than previous general-purpose algorithms using similar quantum resources. 
 Moreover, since the coin is defined by a single observable, the method lends itself well to quantum 
 error mitigation.
We test this in practice with a proof-of-concept 9-qubit experiment, where we successfully mitigate errors through a simple noise-extrapolation procedure.
Our findings offer an interesting alternative for quantum partition function estimation relevant to early-fault quantum hardware. 
\end{abstract}
\maketitle

\section{\label{sec:intro}Introduction}
The need to determine normalization constants for probability distributions or partition functions appears in many areas ranging from statistical physics \cite{huang1987statistical, Mark1993} to molecular biology \cite{FREIRE1994502,zwanzig1997,friedman2004inferring} and generative machine learning \cite{Harshvardhan2020generative}. In machine learning with probabilistic graphical models, computing the partition function is essential for statistical inference, and it is typically the hardest part of the methods \cite{koller_probabilistic2009}.
 For example, for restricted Boltzmann machines, a popular graphical model with a high degree of structure, even approximating it to within a large multiplicative factor is classically hard \cite{Long2010RBM}.
In statistical physics, in turn, knowing the partition function of a system in equilibrium at a fixed temperature allows one to calculate the free energy and important thermodynamic quantities such as magnetization and specific heat. 
Despite its fundamental importance, calculating or even approximately estimating partition functions may involve summing over an exponential number of possible configurations, and it is known to be a classically intractable problem in general. 
For classical Hamiltonians, the exact computation is closely related to counting problems and is, in the worst case, \verb|#|$\P$-hard \cite{Bulatov2005}; while 
partition function estimation (PFE) is in $\BPP^\NP$, in general \cite{Stockmeyer_1983}.
On the other hand, the complexity of multiplicative-error approximate PFE at arbitrary inverse temperature $\beta$ for quantum Hamiltonians is unknown \cite{Bravyi_2022}. 

Naively, to calculate the partition function of a given 
$N$-dimensional quantum Hamiltonian $H$ in a classical computer, one needs $\Tilde{\mathcal{O}}\big(N^\omega\big)$ operations to diagonalize $H$, with $\omega<3$ the exponent of matrix multiplication \cite{banks2023pseudospectral}. 
This scaling is highly unsatisfactory because $N$ itself scales exponentially with the number of particles, typically the relevant figure of merit for the problem's size.
This scaling can be improved for approximate calculations if some underlying problem structure is available. For instance, the kernel polynomial method offers a run-time scaling of $\mathcal{O}(N)$ for sparse matrices \cite{Wei_e_2006}. 
Moreover, the partition function can be approximated efficiently for certain, highly structured Hamiltonian models and parameter regimes \cite{Istrail2000, weitz2006, Sinclair2014, Barvinok_2021, Bravyi_2022}. 
However, an efficient general-purpose classical algorithm for PFE, i.e., that works for any model (classical or quantum) and inverse temperature, is not expected to exist \cite{Alhambra2023}. 
In particular, the problem must display a transition in complexity, from easy to hard, as the inverse temperature increases, since it is trivial for $\beta=0$ but $\QMA$-hard for sufficiently high $\beta$, even for 2-local Hamiltonians \cite{Harrow_2020}.

Several quantum algorithms for PFE have been proposed 
for both classical \cite{Wocjan2009, Montanaro_2015,Harrow_2020_partition,Arunachalam_2022} and quantum \cite{Poulin2009,chowdhury_computing_2021,Bravyi_2022,Jackson_2023,tosta2023randomized} Hamiltonians, culminating in quantum algorithms \cite{Bravyi_2022} with runtime 
$\Tilde{\mathcal{O}}\left(\sqrt{\frac{N}{Z_\beta}}\,\frac{\beta}{\varepsilon_{r}}\right)$ to estimate a partition function $Z_\beta$ up to relative precision $\varepsilon_r$ for general positive Hamiltonians. 
 Although favorable in runtime, these techniques 
 rely on costly subroutines such as quantum phase estimation (QPE) \cite{kitaev1995quantum}, and quantum amplitude estimation (QAE) and amplification (QAA)  \cite{Brassard_2002}. 
 Methods that do not use these subroutines --- hence   
 requiring significantly smaller quantum circuits --- have been recently proposed \cite{chowdhury_computing_2021,tosta2023randomized}. However, these incur in runtimes $\Tilde{\mathcal{O}}\left({\frac{N^2}{Z_\beta^2}}\,\frac{\sqrt{\beta}}{\varepsilon_{r}^2}\right)$, which narrows the regime of potential advantage over classical methods. An overview of the existing quantum algorithms for PFE is shown 
 in Table~\ref{tab:overview}, along with their required quantum resources.

\begin{table*}[ht!]
    \centering
    \scalebox{0.85}{
    \begin{tblr}{
    width=1.1\textwidth,
  colspec = {X[c,m,-1]X[c,m,-1]X[c,m,-1]X[c,m,-1]X[c,m,-1]X[c,m,-1]},
  stretch = 0,
  colsep = 2pt,
  rowsep = 3pt,
  hlines = {0.4pt},
  vlines = {0.4pt},
}
        \thead{Ref.} & \thead{Runtime} & \thead{\# of ancillas} & \thead{Requirements} & \thead{Hamiltonian} & \thead{Access model} \\
         \hline
          \cite{Poulin2009} &  $\Tilde{\mathcal{O}}\left(\sqrt{\frac{N}{Z_\beta}} \frac{\beta^5}{\varepsilon_r^2}\right)$  &  $a+\mathcal{O}\left(n + \beta \log \frac{\beta}{\varepsilon_r}\right) $ & {QPE, QAA, QAE,\\ cooling schedule} & {quantum \\ $H\geq 0$}  & $e^{-itH}$\\
        \cite{Wocjan2009,Montanaro_2015, Harrow_2020_partition, Arunachalam_2022} & $\Tilde{\mathcal{O}}\left(\frac{n\, \sqrt{\tau}}{\varepsilon_r}\right)$ & $a+\mathcal{O}\left(\log\tau\, \log(\frac{1}{\varepsilon_r})\right)$ &  {QAE, QPE, \\ cooling schedule} & {classical\\$H\geq0$} &  {Markov chain quantum walk \\with relaxation time $\tau$} \\
        \cite{zhang2023dissipative} &  {$\mathcal{O}\left(\frac{1}{\varepsilon_r^2\,\chi}\exp[\frac{2\beta\, m}{(1-\chi)^{2m-1}}]\right)$} & $\mathcal{O}(1)$ & -- &  {$m$ local terms $h_j$\\ $\kappa=\sum_{j=1}^m \|h_j\|$}& implementation of each $h_j$\\
        \cite{Bravyi_2022} & \( \mathcal{O}\left(\frac{1}{\varepsilon_r} \sqrt{\frac{N}{Z_\beta}} (\beta + \log\frac{1}{\varepsilon_r})\right) \) & $a+\mathcal{O}\left(\log(\frac{1}{\varepsilon_r})\right)$ & QAE & $H\geq0$ &  {block-encoding of $H'$\\ (effective $\sqrt{H}$ ) }\\
        \cite{rouze2024optimalquantumalgorithmgibbs} & $ \Tilde{\mathcal{O}}\left(\frac{n^3}{\varepsilon_r^2}\right) $ & $a+\mathcal{O}(1)$ & {Cooling schedule,\\ $\beta\leq\frac{1}{615^D J}$, $J=hkl$} & { $k$-local terms $h_j$, $\|h_j\|\leq h$, \\ each qubit in $l'<l$ terms} &  {block-encoding of $H$ }\\
         \cite{chowdhury_computing_2021} & $ \Tilde{\mathcal{O}}\Big( \big(\frac{N}{Z_\beta}\big)^2 \frac{e^{2\beta} 2^{2 a}}{\varepsilon_{\text{r}}^2} (\beta^2 + n^2)\Big)$ & $2a+1$ & -- & any & block-encoding of $H$ \\
         \cite{tosta2023randomized} & $ \Tilde{\mathcal{O}}\Big(\frac{e^{2\beta}N^2}{\varepsilon_{\text{r}}^2 Z_\beta^2} \sqrt{\beta}\Big) $ & $a+1$ & -- & any & block-encoding of $H$\\
        this work & $\Tilde{\mathcal{O}}\left({\frac{N}{Z_\beta}}\,\frac{e^\beta}{\varepsilon_{r}^2}\right)$ & $a$ & -- & any & block-encoding of $e^{-\beta H}$\\
        {this work\\+ QSP} & $\Tilde{\mathcal{O}}\left({\frac{N}{Z_\beta}}\,\frac{e^\beta}{\varepsilon_{r}^2}\,\sqrt{\beta}\right)$ & $a$ & -- & any & block-encoding of $H$ \\
   \end{tblr}
   }
    \caption{{\bf Overview of quantum algorithms for PFE.} Comparison 
    in terms of runtime, 
    number of ancillas, subroutines required, 
    type of Hamiltonian supported, and the access model -- i.e., a quantum oracle 
    that encodes the input $H$ (potentially requiring 
    $a$ ancillary qubits). 
    Runtimes are given in query complexity, i.e., in terms of the total number of calls to the access model
    . The dimension of $H$ is $N=2^n$, where $n$ is the number of system qubits, $\chi=\frac{\varepsilon_\text{r}}{\kappa\, m^2\beta}$ for ref. \cite{zhang2023dissipative}, and $D$ is the lattice dimension for ref. \cite{rouze2024optimalquantumalgorithmgibbs}. In turn, the cooling schedule refers to a classical subroutine yielding a suitable list $\beta_0=0\leq\beta_1\leq\cdots\leq\beta_l=\beta$. The table does not include heuristic approaches based on 
    variational circuits 
    \cite{Wu_2022, Matsumoto_2022}. 
    Among the methods in the table, the dissipative quantum Gibbs sampler of Ref. \cite{zhang2023dissipative} is convenient 
    in terms of the number of ancillas and avoiding costly Hamiltonian oracles and QPE/QAE. 
    However, the number of interaction terms $m$ usually scales polynomially in $n$ [for instance, $ m = \mathcal{O}(n) $ for geometrically local system in regular lattices and $m=\mathcal{O}(n^4)$ for typical molecules in chemistry],
    making this approach's complexity, in general, significantly 
    worse than $\mathcal{O}(2^{\beta n})$.\footnote{\label{firstfootnote}The runtime presented in the table includes the sample complexity, not shown in Ref. \cite{zhang2023dissipative}. However, it is calculated using the same tools we use in Thm. \ref{thm:prob}, yielding to the required number of runs of their algorithm being $\Tilde{\mathcal{O}}\big(\frac{1}{\varepsilon_r^2}\big)$.}    Markov chain methods feature promising runtimes; however, the difficulty of implementing 
    quantum walks and the 
    hardness 
    of estimating 
    $\tau$ 
    make this scheme challenging \cite{Lemieux2020efficientquantum, Wocjan_2023}. Ref.~\cite{rouze2024optimalquantumalgorithmgibbs} presents a rapidly mixing method that is provably efficient for high temperatures (low $\beta$).   
    In the last row, we assume 
    that $e^{-\beta H}$ is implemented via 
    QSP using the best known polynomial approximation, which yields 
    a query complexity 
    $\Tilde{\mathcal{O}}(\sqrt{\beta})$ per block-encoding of $e^{-\beta H}$. 
    Our algorithm features a 
    quadratic improvement in $\frac{N\,e^\beta}{Z_\beta}$ 
    over other methods \cite{chowdhury_computing_2021,tosta2023randomized}
    requiring similar quantum resources.}
    \label{tab:overview}
\end{table*}

\begin{figure}[t!]
    \centering
    \includegraphics[width=1\columnwidth]{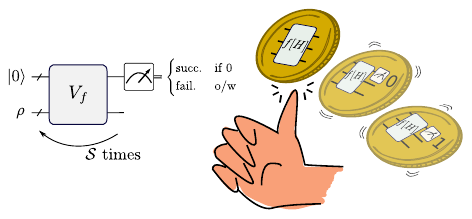}
    \caption{\textbf{Quantum coin.} A quantum-coin toss (relative to an input state $\rho$) consists of first applying a unitary quantum circuit implementing a block-encoding  $V_{f}$ of the Hamiltonian function $f[H]$ on the system and an ancillary register in a reference state $\ket{0}$ and then measuring the ancillary register in the computational basis. If the ancillary measurement returns $0$ (successful post-selection for the correct block of $V_{f}$), we consider the toss result as heads and else as tails (failed post-selection ). The matrix $f[H]$ and the system's initial state $\rho$ determine the probability of these two outcomes. Our quantum algorithms for partition-function estimation are based on $\mathcal{S}$ quantum-coin tosses with the choices $f[H]=f_\beta[H]=e^{-\beta H/2}$ and $\rho$ the maximally mixed state (see Sec. \ref{sec:PFE}).
    }
    \label{fig:coin}
\end{figure}

This work presents a general-purpose quantum algorithm for PFE using a quantum coin as depicted in Fig. \ref{fig:coin}. The coin toss is defined as the result (success or failure) of a measurement on a quantum circuit implementing quantum imaginary-time evolution (QITE). More specifically, 
our algorithm requires implementing a unitary block-encoding of the operator function $f_\beta[H]= e^{-\beta H/2}$ -- which can be realized, for instance, using quantum signal processing (QSP) \cite{LowChuangQuantum2019,Gilyen2019,silva2022fourierbased, Dong_2022}-- followed by a measurement on the block-encoding ancillas to determine if the correct block has been applied to the system (heads) or not (tails).
The partition function can hence be estimated directly from samples of quantum coin toss. This implies that the algorithm does not require any ancillas besides the ones used for block-encoding. We offer two variants of the algorithm, one of which estimates the heads probability and the other one the number of runs between two successive heads. The latter automatically gives a relative-error PFE, while the natural result of most algorithms is an additive precision estimation. 
In both cases, the number of required coin tosses scales as $\Tilde{\mathcal{O}}\left({\frac{N}{Z_\beta}}\,\frac{1}{\varepsilon_{r}^2}\right)$, quadratically better in $\frac{N}{Z_\beta}$  than previous algorithms that also do not use QPE and QAE. The tighter bound for the required number of samples comes from interpreting the probabilistic implementation of non-unitary operator functions as a coin, allowing us to use tools from Bernoulli processes.

We perform 
a proof-of-principle deployment 
of our algorithm on an IonQ quantum processor. We run 
experiments for two different Hamiltonian models, an Ising model on a 4-vertex random graph and the Quantum Restricted Boltzmann Machine (QRBM) \cite{Melko2019} for 2 visible and 2 hidden neurons, and different values of $\beta$ from zero until  
$Z_{\beta}$ converges to its asymptotic value. Since the general operation of the algorithm and the number of coin tosses are independent of how the block-encoding is generated, for simplicity, we resort to a variational implementation where we train a hardware-efficient Ansatz circuit of up to 20 layers to approximate the desired block-encoding unitary. 
Our experiments involve 9 qubits in total: 4 for the system, 4 to obtain a purification of the maximally mixed state, and 1 block-encoding ancilla. By assuming a single-parameter error model, we could dedicate a small portion of the total samples to learning the error model and thus inverting the noise channel of all measurements performed. Surprisingly, besides the simplicity, our results show remarkable agreement with exact calculations, highlighting the experimental friendliness of our approach.

The paper is organized as follows: In Sec. \ref{sec:quantum_coin}, we define the quantum coin in the general case, meaning for any function of a Hamiltonian. We apply the coin to the PFE problem in Sec. \ref{sec:PFE} using two different statistical analyses, the first to estimate the heads probability and the second to estimate the average number of coin tosses between two heads results. In Sec. \ref{sec:exp_imp}, we discuss the experimental implementation of our method. The technical experimental details are presented in Sec. \ref{sec:methods}, along with a discussion about implementing the quantum coin using QSP and the implications of simulated annealing to our method. We conclude the paper with final discussions in Sec. \ref{sec:conclusion}.



\section{The quantum coin}
\label{sec:quantum_coin}

Let $H$ be a Hamiltonian of $n$ qubits. We assume the ability to block-encode a function $f[H]$ into a unitary transformation $V_{f}$ acting on $n+a$ qubits whose corresponding circuit compilation we can access -- for short, we omit the $H$ dependence in $V_f$.
It means that $V_f$ must satisfy 
\begin{equation}\label{eq:qsp}
    \|\bra{0^{\otimes a}}V_f\ket{0^{\otimes a}}-\alpha\,f[H]\|\leq \varepsilon'
\end{equation}
for a given target precision $0<\varepsilon'\leq 1$ and sub-normalization $0<\alpha \leq1$ such that $\|\alpha\,f[H]\|\leq1$, where $\|\cdot\|$ is the spectral norm. Therefore, having initialized the $a$ ancillas in $\ket{0^{\otimes a}}$, the implementation of $f[H]$ up to error $\varepsilon'$ on $n$-qubits via $V_f$ occurs only after post-selecting the ancillas in $\ket{0^{\otimes a}}$. Defining  $\alpha\Tilde{f}[H]:=\bra{0^{\otimes a}}V_f\ket{0^{\otimes a}}$, if the system is initialized in a state $\rho$, the post-selection succeeds with probability
\begin{equation}\label{eq:prob_suc}
    p_{\suc}=\alpha^2 \Tr[\Tilde f[H]\, \rho\,\Tilde f[H]^\dagger].
\end{equation}

Success or failure in post-selecting the ancillas in state $\ket{0^{\otimes a}}$ correspond to heads or tails of a quantumly flipped coin, depicted in Fig. \ref{fig:coin} and defined next. 

\begin{dfn}[Quantum coin] A quantum coin $C(f[H],\alpha,\varepsilon',\rho)$ is an approximate block-encoding circuit satisfying Eq.~\eqref{eq:qsp}  for a function $f[H]$ acting on the state $\ketbra{0^{\otimes a}}{0^{\otimes a}}\otimes \rho $. A coin toss is the result of a computational basis measurement of the ancilla qubits, giving ``heads'' if the measurement returns only zeros and ``tails'' otherwise. The coin probability is given by Eq.\ \eqref{eq:prob_suc}. 
\end{dfn}

Notice that, although we are only interested in sampling from the target probability distribution of successes, successfully applying the block-encoding of a function also yields an approximate preparation of the state $f[H]\,\rho\, f[H]/\Tr[f[H]\,\rho \,f[H]]$, which could be further used as input for other algorithms.

In the next section, we discuss partition function estimation using the above-defined coin with the operator function $f_\beta[H]=e^{-\beta H/2}$.  Assuming the spectrum of $H$ is contained in the interval $[-1,1]$, the normalization constant required to satisfy the block-encoding condition is $\alpha=e^{-\beta/2}$. One can force it by redefining the Hamiltonian $H\rightarrow H/\Lambda$ with a corresponding rescaling $\beta\rightarrow\Lambda\,\beta$ of the inverse temperature, given an upper bound $\Lambda$ for the spectral norm of $H$.

The general operation of the method is independent of how the block-encoding is generated. One of the first proposals for operator function synthesis used a linear combination of unitaries and time evolution to implement $f_\beta[H]$ via the Hubbard-Stratonovich transformation \cite{Chowdhury2017}. Modern methods based on quantum signal processing (QSP) offer Hamiltonian independent circuit primitives and control over the approximation error by implementing polynomial \cite{LowChuangQuantum2019,Gilyen2019} or Fourier \cite{silva2022fourierbased, Dong_2022} approximations of the target function, depending on the access to a block-encoding or a time evolution oracle to the Hamiltonian, respectively. 
One implementation of $V_{f_\beta}$ is discussed in Sec. \ref{Sec:QSP}.

\section{Partition function estimation}
\label{sec:PFE}

The partition function of a Hamiltonian $H$ at inverse temperature $\beta$ is defined as $Z_\beta= \Tr\left[e^{-\beta H}\right]$. We aim to estimate $Z_\beta $ up to relative precision $\varepsilon_r$, i.e to obtain an empirical estimate $\hat{Z}_\beta$ satisfying
\begin{equation}
    |\hat{Z}_\beta-{Z}_\beta|\leq {Z}_\beta \,\varepsilon_r,
\end{equation}
with confidence $1-\delta$, where $\delta$ is the failure probability of the estimation. 
Relative-precision estimation is relevant for $Z_\beta$ since it yields an estimate for the free energy $F_\beta=-(1/\beta)\,\log( Z_\beta)$ within additive precision $\varepsilon_r$. Moreover, it can be deemed more meaningful than its additive-precision counterpart given the large range of values that $Z_\beta$ can take for varying $\beta$. While the complexity of additive precision PFE is well-characterized with efficient quantum algorithms to solve it -- in fact, it is complete for the class of problems efficiently solvable in a quantum computer in the setting of one pure qubit and all other qubits initialized in the maximally mixed state \cite{brandao2008entanglement,chowdhury_computing_2021},-- little is known about the complexity of the relative precision PFE \cite{Bravyi_2022}.


Taking $\rho=\mathds{1}/2^n$ as the maximally mixed state of the $n$-qubits system, the partition function is written as $Z_\beta=2^n\Tr[e^{-\beta H/2}\,\rho\, e^{-\beta H/2}]$.
By comparison with Eq. \eqref{eq:prob_suc}, it is evident that the probability of success $p_{\suc}^{(\beta)}$ of applying $f_\beta[H]$ to the maximally mixed state is proportional to the partition function in the ideal case of $\varepsilon'=0$. More specifically, 
\begin{equation}\label{eq:part_prob}
    Z_\beta = {e^\beta}{2^n \, p_{\suc}^{(\beta)}}.
\end{equation} 
In other words, the partition function $Z_\beta$ determines (and can be estimated from) the probability of the coin $C\big(f_\beta[H],e^{-\beta/2},\varepsilon',\sfrac{\mathds{1}}{2^n}\big)$. The fact that only an approximate block-encoding of $f_\beta[H]$ can be implemented ($\epsilon'\neq 0$) induces a bias in the estimate controlled by the value of $\epsilon'$. The original estimation problem that, in principle, involves a sum over an exponential number of energy levels, translates into a simple Bernoulli process for a single binary variable. In addition, in the same spirit of simulated annealing, the PFE quantum coin also admits to using cooling schedules to improve its cost. We discuss this in detail in Sec. \ref{sec:fragment}.


We consider two random processes involving the quantum coin that yield solutions to the PFE problem. The first one is a Bernoulli process, properly speaking, in which the output (heads or tails) of each coin flip is the random variable. The second one takes the number of times the circuit is run between two successes as the random variable. While the former presents a better dependence on the confidence $\delta$, the second one has the advantage of directly yielding a relative precision estimation. The estimation's complexity manifests in the number of coin tosses required to achieve the desired accuracy since the probability of success is exponentially small in the inverse temperature and the number of qubits.

\subsection{Success-probability estimation}
\label{sec:PFE_succ_prob}


\begin{algorithm}[t]
\SetNoFillComment
\caption{Partition function estimation from the success probability of the coin}\label{alg:one}
\SetKwInOut{Input}{input}\SetKwInOut{Output}{output}
\Input{coin $C\big(f_\beta[H],e^{-\frac{\beta}{2}},\varepsilon',\frac{\mathds{1}}{2^n}\big)$, number of samples~$\mathcal{S}$}
\Output{an estimate $\hat{Z}_\beta$ for the partition function}
$\mathcal{S}_{\suc}\gets 0$ \Comment*{store total $\#$ of successes}
\For{$j\leftarrow 1$ \KwTo  $\mathcal{S}$}{
  prepare $\ketbra{0^{\otimes a}}{0^{\otimes a}} \otimes \frac{\mathds{1}}{2^n}$\;
  run  $C\big(f_\beta[H],e^{-\frac{\beta}{2}},\varepsilon',\frac{\mathds{1}}{2^n}\big)$
  getting outcome $c_j$\;
  \If{$c_j=\,$``heads''}{$\mathcal{S}_{\suc}\leftarrow  \mathcal{S}_{\suc}+1$}
}
$\hat{p}_{\suc}^{(\beta)} \gets \frac{1}{\mathcal{S}+z_{\delta}^2}\left(\mathcal{S}_{\suc}+\frac{z_{\delta}^2}{2}\right)$  \Comment*[r]{AC estimate \cite{Brown2001}}
return $\hat{Z}_\beta\gets (2^ne^\beta)\,\hat{p}_{\suc}^{(\beta)}$
\end{algorithm}

A sequence of $\mathcal{S}$ tosses of the ideal coin $C\big(f_\beta[H],e^{-\beta/2},{\varepsilon'=0},\sfrac{\mathds{1}}{2^n}\big)$ produces a finite sequence of independent random variables $c_1,\, c_2,\cdots, c_{\mathcal{S}}$ all identically distributed according to $\text{Pr}({c_j=1})=p_{\suc}^{(\beta)}$ and $\text{Pr}({c_j=0})=1-p_{\suc}^{(\beta)}$. To estimate the coin probability, one could simply use the proportion of successful events observed. However, it incurs inconsistencies when the true success probability is close to zero \cite{pires2008interval}, which is generally true for $p_{\suc}^{(\beta)}$. Instead, we define an empirical estimate $\hat{p}_{\suc}^{(\beta)}$
using the so-called Agresti-Coull (AC) interval, a confidence interval for binomial proportions proven to be suited for small probabilities  \cite{Brown2001} (see the details in App.~\ref{sec:proof_prob}). Applying Eq. \eqref{eq:part_prob} defines an estimate  $\hat{Z}_\beta$ for the partition function. An approximation error $\varepsilon'$ in the coin implementation induces a bias in $\hat{p}_{\suc}^{(\beta)}$ relatively to the ideal probability $p_{\suc}^{(\beta)}$, and consequently also in the estimate  $\hat{Z}_\beta$. The overall relative error $\varepsilon_r$ is attained with the choice $\varepsilon'\leq \frac{Z_\beta}{6\,e^\beta2^n}\varepsilon_r$.  
 Obviously, $Z_\beta$ is not known before running the algorithm, so 
 we are forced to take the worst-case tolerated error 
 $\varepsilon'=\frac{1}{6\,e^\beta2^n}\varepsilon_r$. That is, the block-encoding approximation error is rescaled by 
 an exponentially small factor. However, the coin implementation cost with quantum signal processing is 
 logarithmic with $\varepsilon'$, so this 
 does not represent a prohibitive overhead.

Algorithm \ref{alg:one} gives a pseudocode for the procedure just described. Its correctness and complexity are formalized in Thm.~\ref{thm:prob} below. We denote by 
$z_{\delta}$  
the quantile of a standard normal distribution at $1-\delta/2$, i.e., $z_{\delta}$
satisfies $\frac{1}{\sqrt{2\pi}}\int_{-\infty}^{z_\delta}e^{-t^2/2}dt=\frac{\delta}{2}$. We note that, to our convenience, $z_{\delta}$ is an extremely slow-growing function of the confidence. For instance, $z_{.05}=1.96$, while $z_{10^{-9}}=6.11$.

 \begin{thm}[PFE using the success probability]\label{thm:prob}
     Given a quantum coin $C\big(f_\beta[H],e^{-\beta/2},\varepsilon',\sfrac{\mathds{1}}{2^n})$, with approximation error $\varepsilon'\leq \frac{Z_\beta}{6\,e^\beta2^n}\varepsilon_r$, the partition function of $H$ at inverse temperature $\beta$ can be estimated up to relative error $\varepsilon_r$ with confidence $1-\delta$ using $\mathcal{S} = 8\,\frac{z_{\delta}^2}{\varepsilon_r^2}\,\frac{2^ne^\beta}{Z_\beta}$ tosses of the coin.
 \end{thm}

In obtaining the result above, an additive precision estimation was artificially transformed into a relative precision one, resulting in a non-practical dependence of the required number of samples $\mathcal{S}$ on the quantity to be estimated and, hence, it cannot be calculated beforehand. 
However, in Ref.\  \cite{chowdhury_computing_2021}, the authors propose a method to obtain a relative precision estimation for the partition function from iterative additive precision ones. To obtain the desired relative precision, one starts with a loose additive estimation with error $\varepsilon=\varepsilon_{\text{r}}Z_{\text{max}}/2$, where $Z_{\text{max}}$ is the maximum possible value of the partition function. The additive precision estimation is run, halving the additive precision in each step until the $R$-th step when the estimate is larger than $Z_{\text{max}}/2^R$. The number of steps is a random variable whose average is $\mathcal{O}(\log(Z_{\text{max}}/Z_\beta))$. If the confidence of each step $r$ is $1-\frac{6}{\pi^2}\frac{\delta}{r^2}$, then the final estimate is $\varepsilon_{\text{r}}Z_\beta$ close to the true value $Z_\beta$ with confidence $1-\delta$.

\subsection{Trials to a success}
\label{sec:PFE_trials}

We now consider a random process in which we are not exactly interested in the output of each circuit run, but we record the number of times $R$ the circuit is implemented between two consecutive success events. In each trial of this process, the random variable $R$ is not bounded and can take any integer value greater than $1$. After $\mathcal{S}_{\suc}$ successes have been observed, a sequence $R_1,\,R_2,\cdots,R_{\mathcal{S}_{\suc}}$ of independent identically distributed random variables is generated.
The mean value of this random variable is related to the partition function as follows:
\begin{lem}\label{lem:part_succ}
    The partition function can be determined from the average number of trials-until-a-success of an ideal coin $C\big(f_\beta[H],e^{-\beta/2},\varepsilon'=0,\sfrac{\mathds{1}}{2^n}\big)$ as $Z_\beta=e^\beta 2^n/\Bar{R}$. 
\end{lem}

\begin{algorithm}[t]
\caption{Partition function estimation from the number of repetitions until a success}\label{alg:two}
\SetKwInOut{Input}{input}\SetKwInOut{Output}{output}
\Input{coin $C\big(f_\beta[H],e^{-\beta/2},\varepsilon',\sfrac{\mathds{1}}{2^n}\big)$, number of successes $\mathcal{S}_{\suc}$}
\Output{an estimate $\hat{Z}_\beta$ for the partition function $\Tr(e^{-\beta H})$}
\For{$j\leftarrow 1$ \KwTo  $\mathcal{S}_{\suc}$}{
  $R_j\gets 0$\Comment*{store $\#$ of runs to $j$-th success}
  $c\gets\,$``tails''\;
  \While{$c\neq\,$``heads''}{
  $R_j\gets R_j+1$\;
  prepare $\rho\otimes\ketbra{0^{\otimes a}}{0^{\otimes a}}$\;
  run $C\big(f_\beta[H],e^{-\beta/2},\varepsilon',\sfrac{\mathds{1}}{2^n}\big)$
  getting outcome $c$\;}  
}
$\Hat{\Bar{R}}\gets \frac{1}{\mathcal{S}_{\suc}}\sum_{j=1}^{\mathcal{S}_{\suc}}R_j$\;
$\hat{Z}_\beta\gets 2^ne^\beta/\Hat{\Bar{R}}$
\end{algorithm}

Lemma \ref{lem:part_succ} allows us to elaborate an algorithm for partition function estimation whose pseudo-code is found in Alg. \ref{alg:two}. As before, an imperfect implementation of the exponential function with approximation error $\varepsilon'$ results in a biased partition function estimate. Again, the choice $\varepsilon'=\frac{1}{6\,e^\beta2^n}\varepsilon_r$, although not optimal, is enough to keep the target relative error below $\varepsilon_r$. The correctness and complexity of the algorithm in terms of uses of $C\big(f_\beta[H],e^{-\beta/2},\varepsilon',\sfrac{\mathds{1}}{2^n}\big)$ are given in the following theorem:
\begin{thm}[PFE using the number of trials to a success]\label{thm:trials}
    Given a quantum coin $C\big(f_\beta[H],e^{-\beta/2},\varepsilon',\sfrac{\mathds{1}}{2^n}\big)$, with approximation error $\varepsilon'=\frac{1}{6\,e^\beta2^n}\varepsilon_r$,  the partition function of $H$ at inverse temperature $\beta$ can be estimated up to relative precision $\varepsilon_r$ and confidence $1-\delta$ by tossing the coin until $S_{\suc}= \frac{1}{\delta\, \varepsilon_r^2}$ successes are observed.  It requires a total of $\mathcal{S}=(2^n e^\beta)/\left(\delta\,\varepsilon_r^2 Z_\beta\right)$ coin tosses on average.
\end{thm}

Notice that the sample complexity of Alg. \ref{alg:two} is inversely proportional to $\delta$, as opposed to the much milder scaling in Alg. \ref{alg:one}, where the dependence on the confidence appears only through $z_{\delta}$.
However, in contrast to the latter, the former directly delivers the desired relative precision estimation. This is seen from that, in order to obtain 
$Z_\beta$ up to relative precision $\varepsilon_r$, one has to run the 
circuit until $S_{\suc}$ successes are observed, where $S_{\suc}$ does not depend on the specific value of $Z_\beta$. What reflects the hardness of the 
estimation problem in question is the complexity of obtaining a success, which is on average $\Bar{R}=\frac{1}{p_{\suc}^{(\beta)}}=\frac{2^n\,e^\beta}{Z_\beta}$ circuit runs. However, this quantity does not need to be known prior to the experiment. Moreover, one can set $\delta$ to a moderate constant, decreasing the required number of samples, and boost the confidence with a small number of repetitions of Alg. \ref{alg:two}.

\section{Experimental implementation}
\label{sec:exp_imp}

\begin{figure}[t!]
    \centering
    \includegraphics[width=1\columnwidth]{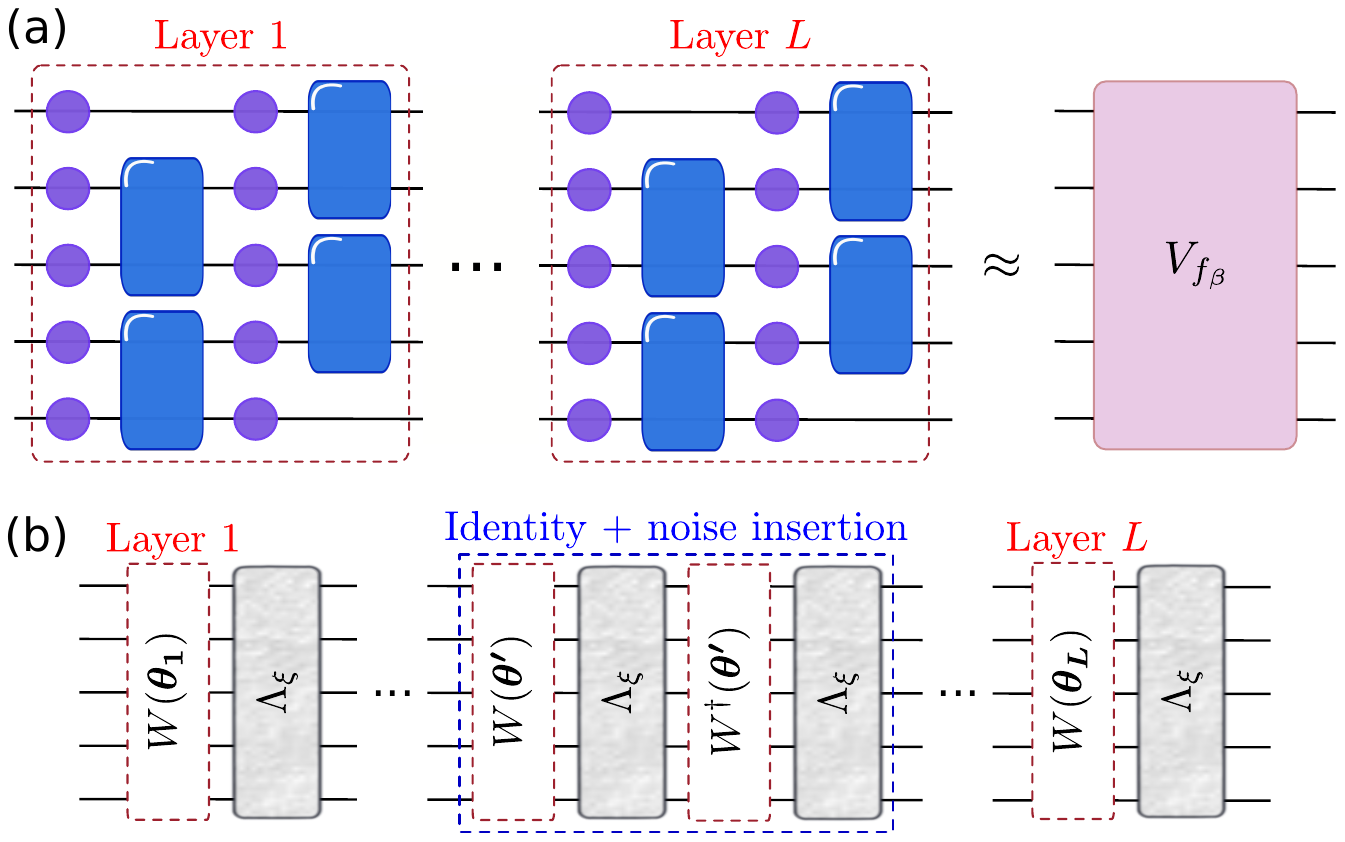}
    \caption{\textbf{Variational quantum circuit.} (a) The ansatz structure implemented in the \textrm{IonQ}'s \texttt{Aria-1} machine. Purple circles represent $\operatorname{GPI2}(\phi)$ gates, and the blue box represent $\operatorname{MS}(\phi_{1}, \phi_{2}, \theta)$ gates. (b) Noise model used for error mitigation. We assume the hardware implements each ideal layer $W(\theta_{j})$, followed by a global depolarizing channel $\Lambda _{\xi}$. Identities in the form of $\mathds{1} = W(\theta')W^{\dagger}(\theta')$ are inserted to increase the noise strength without altering the noiseless result.}
    \label{fig:circuit}
\end{figure}

\begin{figure*}
    \centering
    \includegraphics[width=\textwidth]{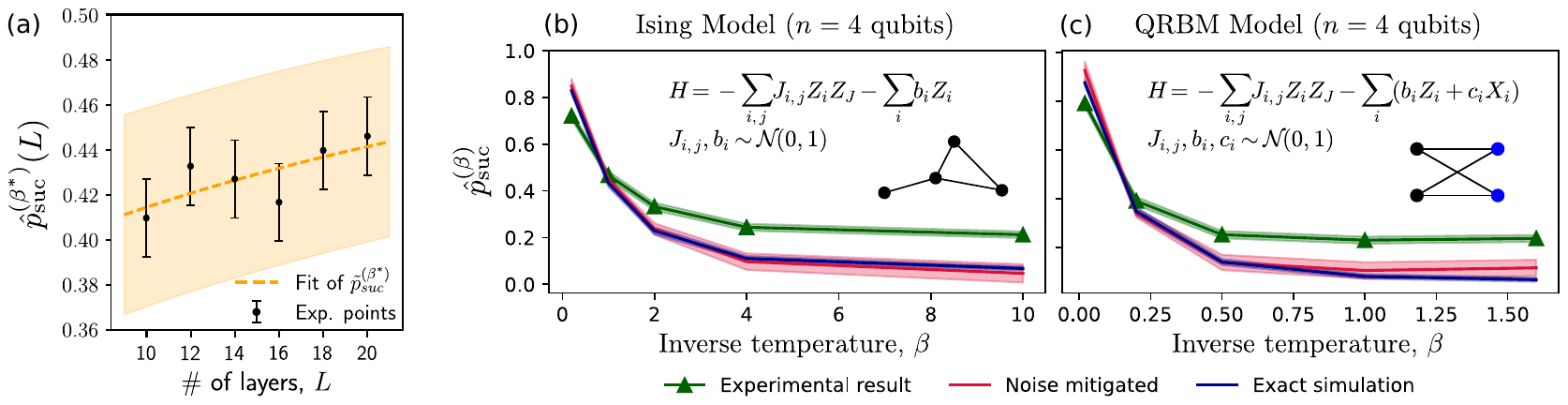}
    \caption{\textbf{Experimental results.} (a) Noise strength determination using identity insertion. The plot shows the empirical value over $3000$ shots of the success probability for different numbers of layers (circuit depths) for fixed $H$ and $\beta^*=0.1$. The dashed line shows the fit of Eq. \eqref{eq:noise_model} which gives $\xi = 0.037 \pm 0.028$ and $\hat{p}^{(\beta^{*})}_{\suc} = 0.38 \pm 0.05$. The uncertainties associated with $\xi$ and $p^{(\beta^{*})}_{\suc}$ are the standard deviations as given by the non-linear least square method of \cite{Vugrin2007}, and determine the $1\sigma$ orange band shown. (b) Probability of success for various values of $\beta$ averaged over $5$ instances of the Ising Model in a random lattice of $n = 4$ qubits. The inset shows the Ising Hamiltonian and an example of a graph instance used. Both the graph and the edges' weights are randomly generated. We explore the low and high (in the sense of $Z_{\beta}$ approaching convergence) $\beta$ regimes. $3000$ measurements are performed on the \textrm{IonQ}'s \texttt{Aria-1} machine for each experimental point. The success probability estimator is the one from Eq.~\eqref{eq:p_suc_estimator}, and its error-mitigated counterpart is obtained from Eq. \eqref{eq:noise_model}. The uncertainty interval for the success probability estimator (green and blue shaded regions) is obtained as in Eq.~\eqref{eq:error_p}. For the noisy mitigated uncertainties (shaded red region), we propagate the uncertainties of $\xi$ and $\hat{p}^{(\beta)}\suc$ through Equation \eqref{eq:noise_model} using standard error propagation formulas \cite{Ku1966}. (c) Same as (b) for the QRBM Model. In the inset, the QRBM Hamiltonian and the graph architecture used, with $2$ visible (black) and $2$ hidden (blue) nodes. We can see an overlap of the error-mitigated curves with the exact ones for both models. This showcases the effectiveness of the simple error mitigation procedure used, as well as the experimental friendliness of our approach.} 
    \label{fig:exp_data}
\end{figure*}

Our algorithm only requires a block-encoding of $e^{-\beta H/2}$, independent of its particular implementation, and does not require any advanced routine such as QPE or QAE, which ultimately allows us to deploy a proof-of-concept experiment on current hardware. 
However, implementing the block-encoding via QSP would be prohibitively costly for the current quantum devices, even for small systems with $n = 4$ qubits, as we consider. Here, we resort to a variational method to optimize the parameters of a layered circuit, as shown in Fig.~\ref{fig:circuit}(a), and obtain a feasible block-encoding unitary circuit instead, with only 1 block-encoding ancilla.  This method is akin to the implementation in Ref.~\cite{Kikuchi_2023}, but we directly block-encode the Hamiltonian function instead of the Hamiltonian. We consider instances of Ising Hamiltonians and quantum restricted Boltzmann machine (QRBM) Hamiltonians with a system size of $4$ qubits (see Fig.~\ref{fig:exp_data}, Sec.~\ref{sec:exp_methods} for details). Four extra qubits are used to obtain a purification of the maximally mixed state from an entangled state of 8 qubits. After the initial state preparation, these qubits are kept idling while the circuit is run, and this step could be substituted by sampling computational basis states uniformly.

Similar to zero-noise extrapolation (ZNE) \cite{temme2017zne,li2017zne}, we 
run the same quantum computation with variable noise levels such that the measurement result at increasing noise can be used to obtain information about the noise model of the hardware. Contrary to standard ZNE, though, where a noise extrapolation is performed for each circuit, we assume a simple noise model that will be learned for one circuit and then used to obtain the noiseless extrapolation of all observables measured in that particular hardware.


We assume that each layer of the brickwork structure implements the ideal noiseless layer unitary  $W(\boldsymbol{\theta})$, followed by an effective global depolarizing noise channel $\Lambda_\xi$ with the same strength $\xi$ for all layers, as shown in Fig.~\ref{fig:circuit}(b). This very simple error model has been recently shown to be a reasonable approximation in experiments in noisy quantum computers \cite{vovorosh2021_mitigation}. Concretely, each layer operation in hardware
is modeled as
\begin{equation}\label{eq:noise_layer}
    \Lambda_{\boldsymbol{\theta}}(\rho):=\xi \frac{\mathds{1}}{2^{n + a}}+(1-\xi) W(\boldsymbol{\theta})\,\rho\, W(\boldsymbol{\theta})^{\dagger},
\end{equation}
where $\xi$ is independent of $\boldsymbol{\theta}$. The ideal circuit unitary is then $V_{f_\beta} = \prod^{L}_{i=1}W(\boldsymbol{\theta}_{i})$, and the overall channel assumed to be implemented by the hardware is the superoperator composition given by $\Lambda_{L}(\rho):= \left(\Lambda_{\boldsymbol{\theta}_{L}}\hdots \,\Lambda_{\boldsymbol{\theta}_{1}}\right)(\rho)$.

According to Eq.~\eqref{eq:part_prob}, the partition function is just the rescaled success probability. Therefore, we refer to both concepts interchangeably. 
Since we are using only $a=1$ ancilla for the block-encoding, Eqs.~\eqref{eq:qsp} and \eqref{eq:prob_suc} require measuring the single qubit observable $P = \ketbra{0^{\otimes a}}{0^{\otimes a}}\otimes\mathds{1}$ whose expectation value gives $p^{(\beta)}_{\suc} = \Tr[V_{f_\beta} \rho V_{f_\beta}^{\dagger} P]$ directly. Using the above map, we can associate the ideal expectation value to the experimentally measured one, i.e.
\begin{align}\label{eq:noise_model}
    \nonumber
    \overline{p}^{(\beta)}_{\suc}(L) &\coloneqq \Tr[\Lambda_{L}(\rho) P] \\
   &= \frac{1 - (1-\xi)^L}{2} + (1-\xi)^L p^{(\beta)}_{\suc}\,.
\end{align}

We use Eq. \eqref{eq:noise_model}, with $\xi = 0.037 \pm 0.028$ obtained from the fitting shown in Fig.~\ref{fig:exp_data}(a) (see Sec. \ref{sec:exp_methods} for details on this procedure), to invert the noise map in all measurements for all circuits considered. The effect of this rather simple error mitigation can be seen in Fig. \ref{fig:exp_data}(b) and \ref{fig:exp_data}(c), where the red lines show the corrected experimental data. Although the standard deviation obtained from the non-linear fit is quite high (upwards of $75\%$ for $\xi$, for instance), the error-mitigated mean values (red curve) are compatible with the exact simulation for most points in the graph for both Hamiltonian models.

\section{Methods}
\label{sec:methods}

\subsection{Coin implementation using QSP}
\label{Sec:QSP}
Let us discuss the cost of one use of $C\big(f_\beta[H],e^{-\beta/2},\varepsilon',\sfrac{\mathds{1}}{2^n}\big)$ when the block-encoding unitary $V_{f_\beta}$ is implemented using quantum signal processing. This enables the total gate costs of Algs. \ref{alg:one} and \ref{alg:two} to be written down.

Consider a unitary oracle $U_H$ acting on $n+b$ qubits and encoding  $H$. Quantum signal processing \cite{LowChuangQuantum2019, Gilyen2019}  offers the recipe to implement functions of $H$ from controlled calls to $U_H$ (and its inverse) interspersed with rotations on the control qubit. In this way, the total number of ancillas is $a=b+\mathcal{O}(1)$ and is determined by the specific oracle model and oracle implementation. Given a function $f[H]$, the recipe produces a circuit on $n+a$ qubits whose resulting unitary transformation $V_f$ satisfies Eq.\ \eqref{eq:qsp}. Two common oracle models are the block-encoding oracle \cite{LowChuangQuantum2019, Gilyen2019} and the time-evolution oracle \cite{silva2022fourierbased, Dong_2022}. The former has $H$ as its upper left block, i.e., $\bra{0^{\otimes b}}U_H\ket{0^{\otimes b}}=H$, and  QSP yields a Chebyshev approximation $\Tilde{f}[H]=\sum_{k=0}^{d}c_k\,T_k[H]$, where $T_k[\cdot]$ is the degree-$k$ Chebyshev polynomial of the first kind.
The latter is the time evolution generated by $H$ at time $t$, i.e. $U_H=e^{-itH}$. In this case, the QSP circuit yields a Fourier approximation $\Tilde{f}[H]=\sum_{k=-d}^d c_k\,e^{i k \omega_0 H}$ of $f[H]$ instead, with $\omega_0$ the fundamental Fourier frequency determined by the choice of $t$. 

The total number of calls to $U_H$ naturally gives the complexity of an algorithm based on QSP. The query complexity ${q}$, defined as the number of times $U_H$ or $U_H^\dagger$ are repeated in the circuit, is proportional to the degree $d$ of the polynomial or Fourier approximation, which ultimately depends on the target precision $\varepsilon'$ and the smoothness of the function $f$. The number of circuit executions $\mathcal{S}$ times the circuit complexity gives the total query complexity $\mathcal{Q}=\mathcal{S}\,{q}$. In turn, for a particular oracle model and implementation with gate complexity $g$, the total gate depth and total runtime are obtained as $g\,q$ and $g\,\mathcal{Q}$, respectively.

Within the block-encoding input oracle model, the QITE propagator $f_\beta[H]$ can be implemented using the Sachdeva-Vishnoi approximation \cite{sachdeva_approximation_2013}, which has degree 
\begin{equation}\label{eq:SachVish}
d=\mathcal{O}(\sqrt{\beta}\log(1/\varepsilon')).    
\end{equation}
 This Chebyshev approximation was proven to be very close to the analytical Jacobi-Anger series \cite{tosta2023randomized}. Therefore, both series give the same truncation order. With $\varepsilon'=\frac{1}{6\,e^\beta2^n}\varepsilon_r$, the coin cost is $q=\mathcal{O}\Big(\sqrt{\beta}\big(\beta+n+\log(1/\varepsilon_r)\big)\Big)$ calls to the Hamiltonian oracle. Finally, from Theorems \ref{thm:prob} and \ref{thm:trials} we obtain $\mathcal{O}\Big(\frac{z_{\delta}^2}{\varepsilon_r^2}\,\frac{2^ne^\beta}{Z_\beta} \sqrt{\beta}\,\big(\beta+n+\log(1/\varepsilon_r)\big)\,g\Big)$ and $\mathcal{O}\Big(\frac{1}{\delta\varepsilon_r^2}\,\frac{2^ne^\beta}{Z_\beta} \sqrt{\beta}\,\big(\beta+n+\log(1/\varepsilon_r)\big)\, g\Big)$ 
for the gate costs of Algs. \ref{alg:one} and \ref{alg:two}, respectively.

\subsection{Fragmented coin implementation}\label{sec:fragment}

In the same spirit of quantum simulated annealing \cite{Somma_2008}, QITE also admits to using cooling schedules to improve their cost, as was shown in Ref. \cite{silva_fragmented_2022}. Here we discuss the implications of this construction to our PFE method. 

We start by defining an inverse temperature schedule $\beta_0=0\leq\beta_1\leq\cdots\leq\beta_l=\beta/2 $ such that 
\begin{equation}
    e^{-\frac{\beta}{2} H}=\prod_{k=1}^l e^{-\Delta_k H},
\end{equation}
with $\Delta_k=(\beta_{k}-\beta_{k-1})/2$. Therefore, QITE for final inverse temperature $\beta/2$ is achieved by successively implementing QITE with an inverse temperature equal to one of the steps $\Delta_k$. Specifically, starting with the initial state $\rho$, we apply $e^{-\Delta_1 H}$ via its block-encoding followed by a measurement on the block-encoding ancillas. If the measurement returned $0$ for all the ancilla qubits, it means that the system state $e^{-\beta_1/2}\rho \,e^{-\beta_1/2}/\Tr\left(e^{-\beta_1}\rho\right)$ has been successfully prepared and we move on to implement $e^{-\Delta_2 H}$. Otherwise, the process is restarted with a new initial state preparation. Given that step $k-1$ was successful, it is easy to show that the probability of success at step $k$ is $p^{(\Delta_k)}_{\suc}=p^{( \beta_k)}_{\suc}/p^{(\beta_{k-1})}_{\suc}=\frac{Z_{\beta_{k}}}{e^{2\,\Delta_k}Z_{\beta_{k-1}}}$, where $p^{( \beta_k)}_{\suc}$ is the probability of successfully implementing the QITE propagator at $\beta_k$ on the initial state $\rho$. Therefore, the probability that all $l$ steps are run  successfully is
\begin{equation}\label{eq:frag_prob}
    p^{(\Delta_1)}_{\suc}p^{(\Delta_2)}_{\suc}\cdots p^{(\Delta_l)}_{\suc}=\frac{Z_{\beta_{l}}}{e^{\beta}Z_{\beta_{0}}}=\frac{Z_{\beta}}{e^{\beta}2^n}.
\end{equation}
This is the same success probability that we had before, with the difference that now the deepest circuit that implements the full inverse temperature QITE is seldom executed.

The corresponding QITE propagator is not exactly implemented in each step. This results in errors in the probability of success of that step and also in the state that is input to the next step. If for all $k$ the success probability of the $k$-th step has a total error $\varepsilon p^{(\Delta_k)}_{\suc}/l $, then we have for the probability of succeeding in all steps $\left(p^{(\Delta_1)}_{\suc} +\varepsilon\, p^{(\Delta_1)}_{\suc}/l\right) \cdots \left(p^{(\Delta_l)}_{\suc} +\varepsilon\, p^{(\Delta_l)}_{\suc}/l\right)=\frac{Z_{\beta}}{e^{\beta}2^n}(1+\varepsilon/l)^l\approx \frac{Z_{\beta}}{e^{\beta}2^n} (1+\varepsilon)$. We also impose that each step has the same probability of success, let's say $p^{(\Delta_k)}_{\suc}\geq 1/2^b$ for all $k\in[l]$ and some $b>0$. This condition is satisfied if the partition functions of successive steps are such that $e^{-2\beta_k}\,Z_{\beta_{k}}\geq\frac{1}{2^b} e^{-2\,\beta_{k-1}}Z_{\beta_{k-1}}$. According to Eq. \eqref{eq:frag_prob}, this implies that the size of the inverse temperature schedule is $l\geq \frac{n+\beta \log{e} -\log{Z_\beta}}{b}$. In Ref. \cite[Eq. (6)]{silva_fragmented_2022} it is shown that the average number of queries to the block-encoding oracle of $H$ to obtain one success (success in all the fragmented steps) is given as $Q=\sum_{j=1}^l \left(\prod_{k=j}^l p^{(\Delta_k)}_{\suc}\right)^{-1}\,q(\Delta\beta_j,\varepsilon'_j)$. In the particular case of an equal-probabilities schedule, it can be bounded as 
\begin{align}
    Q&\leq \max_{j\in [l]}\,q(\Delta\beta_j,\varepsilon'_j)\, \sum_{j=1}^l 2^{j\,b}\\
    & =\max_{j\in [l]}\,q(\Delta\beta_j,\varepsilon'_j)\frac{2^b(2^{lb}-1)}{2^b-1}\\
    &\approx\max_{j\in [l]}\,q(\Delta\beta_j,\varepsilon'_j) \frac{2^b}{2^b-1}\frac{2^ne^\beta}{Z_\beta}.
\end{align}
The successes or failures (no matter in which step they happen) define a quantum coin exactly like we discussed before, and all the analyses of the required number of samples apply; only the cost to implement the coin is different. Each step's approximation error must be made small to counteract the accumulation of errors through the sequence. Nevertheless, the dependence of the query complexity on the approximation error is logarithmic. Moreover, $\Delta\beta_j$ is much smaller than $\beta/2$ making $\max_{j\in [l]}\,q(\Delta\beta_j,\varepsilon'_j)<q(\beta/2,\varepsilon')$ and reducing the cost of a successful event. 

Simulated annealing for PFE of classical Hamiltonians using Markov chains allows for an efficient algorithm to obtain a Chebyshev cooling schedule \cite{Arunachalam_2022}. In the quantum case using QITE, with the condition being on the success probabilities of each inverse temperature step, it is not clear how to determine a schedule that satisfies such strict conditions in practice, let alone determine the optimal schedule. This remains an interesting open problem. Nonetheless, it is numerically observed that even a uniform schedule with all $\Delta\beta_k=\beta/(2l)$ is sufficient to outperform the probabilistic algorithm by orders of magnitude \cite{silva_fragmented_2022}.

\subsection{Experimental implementation details}\label{sec:exp_methods}

\paragraph{Quantum Coin implementation.} 

As mentioned in Section \ref{sec:exp_imp}, a QSP implementation of the block-encoding would be out of the reach for current quantum hardware. For that reason, we use a more hardware-friendly approach by directly learning the block-encoding unitary via a variational method.

The variational ansatz used in this work is one flavor of the so-called hardware-efficient ansatz. Specifically, each layer is composed of a $\operatorname{GPI2}(\phi)$ gate applied to every qubit, followed by $\operatorname{MS}(\phi_1, \phi_2, \theta)$ gates applied first to pairs of qubits $(2j, 2j + 1)$, $j = 0, 1, 2, \dots, \lfloor \frac{n - 1}{2}\rfloor$, another $\operatorname{GPI2}(\phi)$ gate applied to every qubit, then again $\operatorname{MS}(\phi_1, \phi _2, \theta)$ gates applied to pairs $(2j + 1, 2j + 2)$, $j = 0, 1, 2, \dots, \lfloor \frac{n - 2}{2}\rfloor$, i.e, there are $2\text{-qubit}$ gates applied to the pairs of even index qubits, followed by $2\text{-qubit}$ gates applied to the pairs of odd index qubits \cite{IonQ}. This sequence constitutes a single layer. Figure \ref{fig:circuit}(a) shows the ansatz structure. From now on, we refer to the number of layers in a circuit as $L$.

\paragraph{Hamiltonian model.} We consider two types of Hamiltonians, the Ising Model and the Quantum Restricted Boltzmann Machines (QRBM). For each model, we generate $5$ random instances of $4$ qubits, with their corresponding parameters sampled from a normal distribution with mean $0$ and variance $1$. To prepare the enlarged unitary, we use a single extra qubit as an ancillary system to a total circuit size of $5$ qubits in all cases considered. Another $4$ qubits are used to prepare the system's register in the initial state $\frac{\mathds{1}}{2^n}$, but are otherwise unused. Moreover, the spectral error obtained is $\varepsilon ' < 10^{-2}$ for all trained circuits, where $\varepsilon '$ is as in Eq. \eqref{eq:qsp}. The optimization algorithm used in these variational learning was \texttt{COBYLA} from the standard Python library \texttt{Scipy}. In this way, flipping the quantum coin is to measure only the fifth qubit (last qubit in Figure \ref{fig:circuit}) in the computational basis.

In generating the graphs for the Ising Hamiltonians, each node is connected to a randomly selected node that is not yet connected to any other - this step is taken to ensure that no node is left unconnected. Afterward, each possible edge is placed with probability $0.5$. An example instance can be seen in the inset of Figure \ref{fig:exp_data} (a). For each instance we train $5$ different circuits, for $\beta = 0.2, 1, 2, 4, 10$, with $L = 12$, totaling $22$ x $L = 264$ trainable parameters.
In the case of the QRBM Hamiltonians, graphs with $2$ visible and $2$ hidden nodes are considered. We also train $5$ circuits for each instance, this time the inverse temperature chosen were $\beta = 0.02, 0.2, 0.5, 1.0, 1.6$, with $L = 10$, i.e., $22$ x $L = 220$ trainable parameters, since this model was observed to converge faster to the ground-state than the Ising Model. The QRBM geometry can be seen in the inset of Figure \ref{fig:exp_data} (b), where visible nodes are colored black and hidden notes are colored blue.

We deploy all circuits in \textrm{IonQ}'s \texttt{Aria-1} machine. The success probability estimator, $\hat{p}^{(\beta)}_{\suc}$, as defined in Eq. \eqref{eq:p_suc_estimator} is shown in Fig. \ref{fig:exp_data}. For each model, $\hat{p}^{(\beta)}_{\suc}$ is averaged over the $5$ random instances.

\paragraph{Noise model learning.} Both $\xi$ and an estimation $\hat{p}^{(\beta)}_{\suc}$ of $p^{(\beta)}_{\suc}$ in Equation \eqref{eq:noise_model} are unknown. To determine them, we can run circuits that should give the same result but that have different numbers of layers $L$. Since the structure of the circuit is fixed, and only the parameters are changed, under our assumptions, any circuit could be employed. We choose a particular circuit with $L = 10$ as our baseline - specifically, we randomly select one of the QRBM instances with $\beta \coloneqq \beta^{*} = 0.1$. We then repeatedly insert identities of the form $\mathds{1} = W(\theta_{i})W^{\dagger}(\theta_{i})$  after the $j$-th layer, where both $i$ and $j$ are chosen uniformly among the layers of the previous circuit. With this process, we obtain circuits with $L = [10, \,12, \,14, \,16, \,18, \,20]$ - note that each inserted identify is composed of two layers -, all of which, if run in a noiseless machine, would yield precisely the same $p^{(\beta^{*})}_{\suc}$. Finally, after measuring $P$ in the quantum hardware, for $L = [10, \,12, \,14, \,16, \,18, \,20]$, we can perform a non-linear least square \cite{Vugrin2007} fitting to obtain both $\xi$ and $p^{(\beta^{*})}_{\suc}$. This procedure is depicted in Fig. \ref{fig:exp_data}(a). This procedure yields the value $\xi = 0.037 \pm 0.028$, which is the value used for all circuits in this work.

\section{Conclusion}
\label{sec:conclusion}

We presented a quantum algorithm for partition function estimation (PFE) via a quantum coin toss based on quantum imaginary-time evolution (QITE). The method does not require resource-intensive subroutines such as quantum phase estimation or quantum amplitude estimation but relies solely on a block-encoding of the QITE propagator $f_\beta[H]=e^{-\beta H/2}$. The coin-toss approach brings in two main practical advantages. First, it allows us to treat the algorithm as a Bernoulli process, and so prove a runtime scaling in $\sfrac{N\,e^\beta}{Z_\beta}$ quadratically better than in previous Hamiltonian-agnostic algorithms with similar quantum-resource requirements \cite{chowdhury_computing_2021,tosta2023randomized}.
Second, it makes the method directly amenable to standard quantum error mitigation techniques. 
All these features enable a proof-of-concept experimental deployment of our algorithm on IonQ's commercially available device \texttt{Aria-1}. 
There, we study the performance of our algorithm in practice for 4-qubit Hamiltonians using 9 qubits in total (including the ancillas for the block-encoding and the purification of the maximally-mixed input state), successfully mitigating errors through a simple variant of zero-noise extrapolation  \cite{temme2017zne,li2017zne}.

On the specific topic of PFE, our findings offer an interesting alternative (for both classical and quantum Hamiltonians) relevant to early fault-tolerant quantum hardware.
In turn, from a more general perspective, an interesting prospect is to explore further potential use cases of quantum coins relative to matrix functions other than $f_\beta[H]$. For instance,  in Ref. \cite{wang2023fastergroundstateenergy}, in a conceptually different approach, a quantum coin relative to a probability density function of $H$ is used to implement the accept/reject step of a rejection sampling scheme in the context of ground-state energy estimation. However, it is an exciting open question whether other end-user applications may benefit (in resource scaling or experimental practicality) from single-observable approaches like ours.

\begin{acknowledgments}
The authors thank Anirban Chowdhury, Toby Cubitt, and Alessandro Falco for helpful discussions; and Daniel Stilck França for insightful comments on our manuscript. 
\end{acknowledgments}

\bibliography{ref}

\begin{thebibliography}{55}%
\makeatletter
\providecommand \@ifxundefined [1]{%
 \@ifx{#1\undefined}
}%
\providecommand \@ifnum [1]{%
 \ifnum #1\expandafter \@firstoftwo
 \else \expandafter \@secondoftwo
 \fi
}%
\providecommand \@ifx [1]{%
 \ifx #1\expandafter \@firstoftwo
 \else \expandafter \@secondoftwo
 \fi
}%
\providecommand \natexlab [1]{#1}%
\providecommand \enquote  [1]{``#1''}%
\providecommand \bibnamefont  [1]{#1}%
\providecommand \bibfnamefont [1]{#1}%
\providecommand \citenamefont [1]{#1}%
\providecommand \href@noop [0]{\@secondoftwo}%
\providecommand \href [0]{\begingroup \@sanitize@url \@href}%
\providecommand \@href[1]{\@@startlink{#1}\@@href}%
\providecommand \@@href[1]{\endgroup#1\@@endlink}%
\providecommand \@sanitize@url [0]{\catcode `\\12\catcode `\$12\catcode `\&12\catcode `\#12\catcode `\^12\catcode `\_12\catcode `\%12\relax}%
\providecommand \@@startlink[1]{}%
\providecommand \@@endlink[0]{}%
\providecommand \url  [0]{\begingroup\@sanitize@url \@url }%
\providecommand \@url [1]{\endgroup\@href {#1}{\urlprefix }}%
\providecommand \urlprefix  [0]{URL }%
\providecommand \Eprint [0]{\href }%
\providecommand \doibase [0]{https://doi.org/}%
\providecommand \selectlanguage [0]{\@gobble}%
\providecommand \bibinfo  [0]{\@secondoftwo}%
\providecommand \bibfield  [0]{\@secondoftwo}%
\providecommand \translation [1]{[#1]}%
\providecommand \BibitemOpen [0]{}%
\providecommand \bibitemStop [0]{}%
\providecommand \bibitemNoStop [0]{.\EOS\space}%
\providecommand \EOS [0]{\spacefactor3000\relax}%
\providecommand \BibitemShut  [1]{\csname bibitem#1\endcsname}%
\let\auto@bib@innerbib\@empty
\bibitem [{\citenamefont {Huang}(1987)}]{huang1987statistical}%
  \BibitemOpen
  \bibfield  {author} {\bibinfo {author} {\bibfnamefont {K.}~\bibnamefont {Huang}},\ }\href@noop {} {\emph {\bibinfo {title} {Statistical Mechanics, 2nd edition}}}\ (\bibinfo  {publisher} {Wiley},\ \bibinfo {year} {1987})\BibitemShut {NoStop}%
\bibitem [{\citenamefont {Jerrum}\ and\ \citenamefont {Sinclair}(1993)}]{Mark1993}%
  \BibitemOpen
  \bibfield  {author} {\bibinfo {author} {\bibfnamefont {M.}~\bibnamefont {Jerrum}}\ and\ \bibinfo {author} {\bibfnamefont {A.}~\bibnamefont {Sinclair}},\ }\bibfield  {title} {\bibinfo {title} {Polynomial-time approximation algorithms for the ising model},\ }\href {https://doi.org/10.1137/0222066} {\bibfield  {journal} {\bibinfo  {journal} {SIAM Journal on Computing}\ }\textbf {\bibinfo {volume} {22}},\ \bibinfo {pages} {1087} (\bibinfo {year} {1993})},\ \Eprint {https://arxiv.org/abs/https://doi.org/10.1137/0222066} {https://doi.org/10.1137/0222066} \BibitemShut {NoStop}%
\bibitem [{\citenamefont {Freire}(1994)}]{FREIRE1994502}%
  \BibitemOpen
  \bibfield  {author} {\bibinfo {author} {\bibfnamefont {E.}~\bibnamefont {Freire}},\ }\bibfield  {title} {\bibinfo {title} {Statistical thermodynamic analysis of differential scanning calorimetry data: Structural deconvolution of heat capacity function of proteins},\ }in\ \href {https://doi.org/https://doi.org/10.1016/S0076-6879(94)40062-8} {\emph {\bibinfo {booktitle} {Part B: Numerical Computer Methods}}},\ \bibinfo {series} {Methods in Enzymology}, Vol.\ \bibinfo {volume} {240}\ (\bibinfo  {publisher} {Academic Press},\ \bibinfo {year} {1994})\ p.\ \bibinfo {pages} {502}\BibitemShut {NoStop}%
\bibitem [{\citenamefont {Zwanzig}(1997)}]{zwanzig1997}%
  \BibitemOpen
  \bibfield  {author} {\bibinfo {author} {\bibfnamefont {R.}~\bibnamefont {Zwanzig}},\ }\bibfield  {title} {\bibinfo {title} {Two-state models of protein folding kinetics},\ }\href {https://doi.org/10.1073/pnas.94.1.148} {\bibfield  {journal} {\bibinfo  {journal} {Proc Natl Acad Sci U S A}\ }\textbf {\bibinfo {volume} {94}},\ \bibinfo {pages} {148} (\bibinfo {year} {1997})}\BibitemShut {NoStop}%
\bibitem [{\citenamefont {Friedman}(2004)}]{friedman2004inferring}%
  \BibitemOpen
  \bibfield  {author} {\bibinfo {author} {\bibfnamefont {N.}~\bibnamefont {Friedman}},\ }\bibfield  {title} {\bibinfo {title} {Inferring cellular networks using probabilistic graphical models},\ }\href@noop {} {\bibfield  {journal} {\bibinfo  {journal} {Science}\ }\textbf {\bibinfo {volume} {303}},\ \bibinfo {pages} {799} (\bibinfo {year} {2004})}\BibitemShut {NoStop}%
\bibitem [{\citenamefont {GM}\ \emph {et~al.}(2020)\citenamefont {GM}, \citenamefont {Gourisaria}, \citenamefont {Pandey},\ and\ \citenamefont {Rautaray}}]{Harshvardhan2020generative}%
  \BibitemOpen
  \bibfield  {author} {\bibinfo {author} {\bibfnamefont {H.}~\bibnamefont {GM}}, \bibinfo {author} {\bibfnamefont {M.~K.}\ \bibnamefont {Gourisaria}}, \bibinfo {author} {\bibfnamefont {M.}~\bibnamefont {Pandey}},\ and\ \bibinfo {author} {\bibfnamefont {S.~S.}\ \bibnamefont {Rautaray}},\ }\bibfield  {title} {\bibinfo {title} {A comprehensive survey and analysis of generative models in machine learning},\ }\href {https://doi.org/https://doi.org/10.1016/j.cosrev.2020.100285} {\bibfield  {journal} {\bibinfo  {journal} {Computer Science Review}\ }\textbf {\bibinfo {volume} {38}},\ \bibinfo {pages} {100285} (\bibinfo {year} {2020})}\BibitemShut {NoStop}%
\bibitem [{\citenamefont {Koller}\ and\ \citenamefont {Friedman}(2009)}]{koller_probabilistic2009}%
  \BibitemOpen
  \bibfield  {author} {\bibinfo {author} {\bibfnamefont {D.}~\bibnamefont {Koller}}\ and\ \bibinfo {author} {\bibfnamefont {N.}~\bibnamefont {Friedman}},\ }\href@noop {} {\emph {\bibinfo {title} {Probabilistic Graphical Models: Principles and Techniques - Adaptive Computation and Machine Learning}}}\ (\bibinfo  {publisher} {The MIT Press},\ \bibinfo {year} {2009})\BibitemShut {NoStop}%
\bibitem [{\citenamefont {Long}\ and\ \citenamefont {Servedio}(2010)}]{Long2010RBM}%
  \BibitemOpen
  \bibfield  {author} {\bibinfo {author} {\bibfnamefont {P.~M.}\ \bibnamefont {Long}}\ and\ \bibinfo {author} {\bibfnamefont {R.~A.}\ \bibnamefont {Servedio}},\ }\bibfield  {title} {\bibinfo {title} {Restricted boltzmann machines are hard to approximately evaluate or simulate},\ }in\ \href@noop {} {\emph {\bibinfo {booktitle} {Proceedings of the 27th International Conference on International Conference on Machine Learning}}},\ \bibinfo {series and number} {ICML'10}\ (\bibinfo  {publisher} {Omnipress},\ \bibinfo {address} {Madison, WI, USA},\ \bibinfo {year} {2010})\ p.\ \bibinfo {pages} {703–710}\BibitemShut {NoStop}%
\bibitem [{\citenamefont {Bulatov}\ and\ \citenamefont {Grohe}(2005)}]{Bulatov2005}%
  \BibitemOpen
  \bibfield  {author} {\bibinfo {author} {\bibfnamefont {A.}~\bibnamefont {Bulatov}}\ and\ \bibinfo {author} {\bibfnamefont {M.}~\bibnamefont {Grohe}},\ }\bibfield  {title} {\bibinfo {title} {The complexity of partition functions},\ }\href {https://doi.org/https://doi.org/10.1016/j.tcs.2005.09.011} {\bibfield  {journal} {\bibinfo  {journal} {Theoretical Computer Science}\ }\textbf {\bibinfo {volume} {348}},\ \bibinfo {pages} {148} (\bibinfo {year} {2005})}\BibitemShut {NoStop}%
\bibitem [{\citenamefont {Stockmeyer}(1983)}]{Stockmeyer_1983}%
  \BibitemOpen
  \bibfield  {author} {\bibinfo {author} {\bibfnamefont {L.}~\bibnamefont {Stockmeyer}},\ }\bibfield  {title} {\bibinfo {title} {The complexity of approximate counting},\ }in\ \href {https://doi.org/10.1145/800061.808740} {\emph {\bibinfo {booktitle} {Proceedings of the Fifteenth Annual ACM Symposium on Theory of Computing}}},\ \bibinfo {series and number} {STOC '83}\ (\bibinfo  {publisher} {Association for Computing Machinery},\ \bibinfo {year} {1983})\ p.\ \bibinfo {pages} {118}\BibitemShut {NoStop}%
\bibitem [{\citenamefont {Bravyi}\ \emph {et~al.}(2022)\citenamefont {Bravyi}, \citenamefont {Chowdhury}, \citenamefont {Gosset},\ and\ \citenamefont {Wocjan}}]{Bravyi_2022}%
  \BibitemOpen
  \bibfield  {author} {\bibinfo {author} {\bibfnamefont {S.}~\bibnamefont {Bravyi}}, \bibinfo {author} {\bibfnamefont {A.}~\bibnamefont {Chowdhury}}, \bibinfo {author} {\bibfnamefont {D.}~\bibnamefont {Gosset}},\ and\ \bibinfo {author} {\bibfnamefont {P.}~\bibnamefont {Wocjan}},\ }\bibfield  {title} {\bibinfo {title} {Quantum hamiltonian complexity in thermal equilibrium},\ }\href {https://doi.org/10.1038/s41567-022-01742-5} {\bibfield  {journal} {\bibinfo  {journal} {Nature Physics}\ }\textbf {\bibinfo {volume} {18}},\ \bibinfo {pages} {1367–1370} (\bibinfo {year} {2022})}\BibitemShut {NoStop}%
\bibitem [{\citenamefont {Banks}\ \emph {et~al.}(2023)\citenamefont {Banks}, \citenamefont {Garza-Vargas}, \citenamefont {Kulkarni},\ and\ \citenamefont {Srivastava}}]{banks2023pseudospectral}%
  \BibitemOpen
  \bibfield  {author} {\bibinfo {author} {\bibfnamefont {J.}~\bibnamefont {Banks}}, \bibinfo {author} {\bibfnamefont {J.}~\bibnamefont {Garza-Vargas}}, \bibinfo {author} {\bibfnamefont {A.}~\bibnamefont {Kulkarni}},\ and\ \bibinfo {author} {\bibfnamefont {N.}~\bibnamefont {Srivastava}},\ }\bibfield  {title} {\bibinfo {title} {Pseudospectral shattering, the sign function, and diagonalization in nearly matrix multiplication time},\ }\href@noop {} {\bibfield  {journal} {\bibinfo  {journal} {Foundations of computational mathematics}\ }\textbf {\bibinfo {volume} {23}},\ \bibinfo {pages} {1959} (\bibinfo {year} {2023})}\BibitemShut {NoStop}%
\bibitem [{\citenamefont {Weiße}\ \emph {et~al.}(2006)\citenamefont {Weiße}, \citenamefont {Wellein}, \citenamefont {Alvermann},\ and\ \citenamefont {Fehske}}]{Wei_e_2006}%
  \BibitemOpen
  \bibfield  {author} {\bibinfo {author} {\bibfnamefont {A.}~\bibnamefont {Weiße}}, \bibinfo {author} {\bibfnamefont {G.}~\bibnamefont {Wellein}}, \bibinfo {author} {\bibfnamefont {A.}~\bibnamefont {Alvermann}},\ and\ \bibinfo {author} {\bibfnamefont {H.}~\bibnamefont {Fehske}},\ }\bibfield  {title} {\bibinfo {title} {The kernel polynomial method},\ }\href {https://doi.org/10.1103/revmodphys.78.275} {\bibfield  {journal} {\bibinfo  {journal} {Reviews of Modern Physics}\ }\textbf {\bibinfo {volume} {78}},\ \bibinfo {pages} {275–306} (\bibinfo {year} {2006})}\BibitemShut {NoStop}%
\bibitem [{\citenamefont {Istrail}(2000)}]{Istrail2000}%
  \BibitemOpen
  \bibfield  {author} {\bibinfo {author} {\bibfnamefont {S.}~\bibnamefont {Istrail}},\ }\bibfield  {title} {\bibinfo {title} {Statistical mechanics, three-dimensionality and np-completeness: I. universality of intracatability for the partition function of the ising model across non-planar surfaces (extended abstract)},\ }in\ \href {https://doi.org/10.1145/335305.335316} {\emph {\bibinfo {booktitle} {Proceedings of the Thirty-Second Annual ACM Symposium on Theory of Computing}}},\ \bibinfo {series and number} {STOC '00}\ (\bibinfo  {publisher} {Association for Computing Machinery},\ \bibinfo {address} {New York, NY, USA},\ \bibinfo {year} {2000})\ p.\ \bibinfo {pages} {87–96}\BibitemShut {NoStop}%
\bibitem [{\citenamefont {Weitz}(2006)}]{weitz2006}%
  \BibitemOpen
  \bibfield  {author} {\bibinfo {author} {\bibfnamefont {D.}~\bibnamefont {Weitz}},\ }\bibfield  {title} {\bibinfo {title} {Counting independent sets up to the tree threshold},\ }in\ \href {https://doi.org/10.1145/1132516.1132538} {\emph {\bibinfo {booktitle} {Proceedings of the Thirty-Eighth Annual ACM Symposium on Theory of Computing}}},\ \bibinfo {series and number} {STOC '06}\ (\bibinfo  {publisher} {Association for Computing Machinery},\ \bibinfo {address} {New York, NY, USA},\ \bibinfo {year} {2006})\ p.\ \bibinfo {pages} {140–149}\BibitemShut {NoStop}%
\bibitem [{\citenamefont {Sinclair}\ \emph {et~al.}(2014)\citenamefont {Sinclair}, \citenamefont {Srivastava},\ and\ \citenamefont {Thurley}}]{Sinclair2014}%
  \BibitemOpen
  \bibfield  {author} {\bibinfo {author} {\bibfnamefont {A.}~\bibnamefont {Sinclair}}, \bibinfo {author} {\bibfnamefont {P.}~\bibnamefont {Srivastava}},\ and\ \bibinfo {author} {\bibfnamefont {M.}~\bibnamefont {Thurley}},\ }\bibfield  {title} {\bibinfo {title} {Approximation algorithms for two-state anti-ferromagnetic spin systems on bounded degree graphs},\ }\href {https://doi.org/10.1007/s10955-014-0947-5} {\bibfield  {journal} {\bibinfo  {journal} {Journal of Statistical Physics}\ }\textbf {\bibinfo {volume} {155}} (\bibinfo {year} {2014})}\BibitemShut {NoStop}%
\bibitem [{\citenamefont {Barvinok}\ and\ \citenamefont {Barvinok}(2021)}]{Barvinok_2021}%
  \BibitemOpen
  \bibfield  {author} {\bibinfo {author} {\bibfnamefont {A.}~\bibnamefont {Barvinok}}\ and\ \bibinfo {author} {\bibfnamefont {N.}~\bibnamefont {Barvinok}},\ }\bibfield  {title} {\bibinfo {title} {More on zeros and approximation of the ising partition function},\ }\href {https://doi.org/10.1017/fms.2021.40} {\bibfield  {journal} {\bibinfo  {journal} {Forum of Mathematics, Sigma}\ }\textbf {\bibinfo {volume} {9}},\ \bibinfo {pages} {e46} (\bibinfo {year} {2021})}\BibitemShut {NoStop}%
\bibitem [{\citenamefont {Alhambra}(2023)}]{Alhambra2023}%
  \BibitemOpen
  \bibfield  {author} {\bibinfo {author} {\bibfnamefont {A.~M.}\ \bibnamefont {Alhambra}},\ }\bibfield  {title} {\bibinfo {title} {Quantum many-body systems in thermal equilibrium},\ }\href {https://doi.org/10.1103/PRXQuantum.4.040201} {\bibfield  {journal} {\bibinfo  {journal} {PRX Quantum}\ }\textbf {\bibinfo {volume} {4}},\ \bibinfo {pages} {040201} (\bibinfo {year} {2023})}\BibitemShut {NoStop}%
\bibitem [{\citenamefont {Harrow}\ \emph {et~al.}(2020)\citenamefont {Harrow}, \citenamefont {Mehraban},\ and\ \citenamefont {Soleimanifar}}]{Harrow_2020}%
  \BibitemOpen
  \bibfield  {author} {\bibinfo {author} {\bibfnamefont {A.~W.}\ \bibnamefont {Harrow}}, \bibinfo {author} {\bibfnamefont {S.}~\bibnamefont {Mehraban}},\ and\ \bibinfo {author} {\bibfnamefont {M.}~\bibnamefont {Soleimanifar}},\ }\bibfield  {title} {\bibinfo {title} {Classical algorithms, correlation decay, and complex zeros of partition functions of quantum many-body systems},\ }in\ \href {https://doi.org/10.1145/3357713.3384322} {\emph {\bibinfo {booktitle} {Proceedings of the 52nd Annual ACM SIGACT Symposium on Theory of Computing}}},\ \bibinfo {series and number} {STOC ’20}\ (\bibinfo  {publisher} {ACM},\ \bibinfo {year} {2020})\BibitemShut {NoStop}%
\bibitem [{\citenamefont {Wocjan}\ \emph {et~al.}(2009)\citenamefont {Wocjan}, \citenamefont {Chiang}, \citenamefont {Nagaj},\ and\ \citenamefont {Abeyesinghe}}]{Wocjan2009}%
  \BibitemOpen
  \bibfield  {author} {\bibinfo {author} {\bibfnamefont {P.}~\bibnamefont {Wocjan}}, \bibinfo {author} {\bibfnamefont {C.-F.}\ \bibnamefont {Chiang}}, \bibinfo {author} {\bibfnamefont {D.}~\bibnamefont {Nagaj}},\ and\ \bibinfo {author} {\bibfnamefont {A.}~\bibnamefont {Abeyesinghe}},\ }\bibfield  {title} {\bibinfo {title} {Quantum algorithm for approximating partition functions},\ }\href {https://doi.org/10.1103/PhysRevA.80.022340} {\bibfield  {journal} {\bibinfo  {journal} {Phys. Rev. A}\ }\textbf {\bibinfo {volume} {80}},\ \bibinfo {pages} {022340} (\bibinfo {year} {2009})}\BibitemShut {NoStop}%
\bibitem [{\citenamefont {Montanaro}(2015)}]{Montanaro_2015}%
  \BibitemOpen
  \bibfield  {author} {\bibinfo {author} {\bibfnamefont {A.}~\bibnamefont {Montanaro}},\ }\bibfield  {title} {\bibinfo {title} {Quantum speedup of monte carlo methods},\ }\href {https://doi.org/10.1098/rspa.2015.0301} {\bibfield  {journal} {\bibinfo  {journal} {Proceedings of the Royal Society A: Mathematical, Physical and Engineering Sciences}\ }\textbf {\bibinfo {volume} {471}},\ \bibinfo {pages} {20150301} (\bibinfo {year} {2015})}\BibitemShut {NoStop}%
\bibitem [{\citenamefont {Harrow}\ and\ \citenamefont {Wei}(2020)}]{Harrow_2020_partition}%
  \BibitemOpen
  \bibfield  {author} {\bibinfo {author} {\bibfnamefont {A.~W.}\ \bibnamefont {Harrow}}\ and\ \bibinfo {author} {\bibfnamefont {A.~Y.}\ \bibnamefont {Wei}},\ }\bibinfo {title} {Adaptive quantum simulated annealing for bayesian inference and estimating partition functions},\ in\ \href {https://doi.org/10.1137/1.9781611975994.12} {\emph {\bibinfo {booktitle} {Proceedings of the Fourteenth Annual ACM-SIAM Symposium on Discrete Algorithms}}}\ (\bibinfo  {publisher} {Society for Industrial and Applied Mathematics},\ \bibinfo {year} {2020})\ p.\ \bibinfo {pages} {193–212}\BibitemShut {NoStop}%
\bibitem [{\citenamefont {Arunachalam}\ \emph {et~al.}(2022)\citenamefont {Arunachalam}, \citenamefont {Havlicek}, \citenamefont {Nannicini}, \citenamefont {Temme},\ and\ \citenamefont {Wocjan}}]{Arunachalam_2022}%
  \BibitemOpen
  \bibfield  {author} {\bibinfo {author} {\bibfnamefont {S.}~\bibnamefont {Arunachalam}}, \bibinfo {author} {\bibfnamefont {V.}~\bibnamefont {Havlicek}}, \bibinfo {author} {\bibfnamefont {G.}~\bibnamefont {Nannicini}}, \bibinfo {author} {\bibfnamefont {K.}~\bibnamefont {Temme}},\ and\ \bibinfo {author} {\bibfnamefont {P.}~\bibnamefont {Wocjan}},\ }\bibfield  {title} {\bibinfo {title} {Simpler (classical) and faster (quantum) algorithms for gibbs partition functions},\ }\href {https://doi.org/10.22331/q-2022-09-01-789} {\bibfield  {journal} {\bibinfo  {journal} {Quantum}\ }\textbf {\bibinfo {volume} {6}},\ \bibinfo {pages} {789} (\bibinfo {year} {2022})}\BibitemShut {NoStop}%
\bibitem [{\citenamefont {Poulin}\ and\ \citenamefont {Wocjan}(2009)}]{Poulin2009}%
  \BibitemOpen
  \bibfield  {author} {\bibinfo {author} {\bibfnamefont {D.}~\bibnamefont {Poulin}}\ and\ \bibinfo {author} {\bibfnamefont {P.}~\bibnamefont {Wocjan}},\ }\bibfield  {title} {\bibinfo {title} {Sampling from the thermal quantum gibbs state and evaluating partition functions with a quantum computer},\ }\href {https://doi.org/10.1103/PhysRevLett.103.220502} {\bibfield  {journal} {\bibinfo  {journal} {Phys. Rev. Lett.}\ }\textbf {\bibinfo {volume} {103}},\ \bibinfo {pages} {220502} (\bibinfo {year} {2009})}\BibitemShut {NoStop}%
\bibitem [{\citenamefont {Chowdhury}\ \emph {et~al.}(2021)\citenamefont {Chowdhury}, \citenamefont {Somma},\ and\ \citenamefont {Subasi}}]{chowdhury_computing_2021}%
  \BibitemOpen
  \bibfield  {author} {\bibinfo {author} {\bibfnamefont {A.~N.}\ \bibnamefont {Chowdhury}}, \bibinfo {author} {\bibfnamefont {R.~D.}\ \bibnamefont {Somma}},\ and\ \bibinfo {author} {\bibfnamefont {Y.}~\bibnamefont {Subasi}},\ }\bibfield  {title} {\bibinfo {title} {Computing partition functions in the one clean qubit model},\ }\href {https://doi.org/10.1103/PhysRevA.103.032422} {\bibfield  {journal} {\bibinfo  {journal} {Physical Review A}\ }\textbf {\bibinfo {volume} {103}},\ \bibinfo {pages} {032422} (\bibinfo {year} {2021})},\ \bibinfo {note} {arXiv:1910.11842 [quant-ph]}\BibitemShut {NoStop}%
\bibitem [{\citenamefont {Jackson}\ \emph {et~al.}(2023)\citenamefont {Jackson}, \citenamefont {Kapourniotis},\ and\ \citenamefont {Datta}}]{Jackson_2023}%
  \BibitemOpen
  \bibfield  {author} {\bibinfo {author} {\bibfnamefont {A.}~\bibnamefont {Jackson}}, \bibinfo {author} {\bibfnamefont {T.}~\bibnamefont {Kapourniotis}},\ and\ \bibinfo {author} {\bibfnamefont {A.}~\bibnamefont {Datta}},\ }\bibfield  {title} {\bibinfo {title} {Partition-function estimation: Quantum and quantum-inspired algorithms},\ }\bibfield  {journal} {\bibinfo  {journal} {Physical Review A}\ }\textbf {\bibinfo {volume} {107}},\ \href {https://doi.org/10.1103/physreva.107.012421} {10.1103/physreva.107.012421} (\bibinfo {year} {2023})\BibitemShut {NoStop}%
\bibitem [{\citenamefont {Tosta}\ \emph {et~al.}(2024)\citenamefont {Tosta}, \citenamefont {de~Lima~Silva}, \citenamefont {Camilo},\ and\ \citenamefont {Aolita}}]{tosta2023randomized}%
  \BibitemOpen
  \bibfield  {author} {\bibinfo {author} {\bibfnamefont {A.}~\bibnamefont {Tosta}}, \bibinfo {author} {\bibfnamefont {T.}~\bibnamefont {de~Lima~Silva}}, \bibinfo {author} {\bibfnamefont {G.}~\bibnamefont {Camilo}},\ and\ \bibinfo {author} {\bibfnamefont {L.}~\bibnamefont {Aolita}},\ }\bibfield  {title} {\bibinfo {title} {Randomized semi-quantum matrix processing},\ }\href {https://doi.org/10.1038/s41534-024-00883-0} {\bibfield  {journal} {\bibinfo  {journal} {npj Quantum Inf}\ }\textbf {\bibinfo {volume} {10}},\ \bibinfo {pages} {93} (\bibinfo {year} {2024})}\BibitemShut {NoStop}%
\bibitem [{\citenamefont {Kitaev}(1995)}]{kitaev1995quantum}%
  \BibitemOpen
  \bibfield  {author} {\bibinfo {author} {\bibfnamefont {A.~Y.}\ \bibnamefont {Kitaev}},\ }\href@noop {} {\bibinfo {title} {Quantum measurements and the abelian stabilizer problem}} (\bibinfo {year} {1995}),\ \Eprint {https://arxiv.org/abs/quant-ph/9511026} {arXiv:quant-ph/9511026} \BibitemShut {NoStop}%
\bibitem [{\citenamefont {Brassard}\ \emph {et~al.}(2002)\citenamefont {Brassard}, \citenamefont {Høyer}, \citenamefont {Mosca},\ and\ \citenamefont {Tapp}}]{Brassard_2002}%
  \BibitemOpen
  \bibfield  {author} {\bibinfo {author} {\bibfnamefont {G.}~\bibnamefont {Brassard}}, \bibinfo {author} {\bibfnamefont {P.}~\bibnamefont {Høyer}}, \bibinfo {author} {\bibfnamefont {M.}~\bibnamefont {Mosca}},\ and\ \bibinfo {author} {\bibfnamefont {A.}~\bibnamefont {Tapp}},\ }\bibfield  {title} {\bibinfo {title} {Quantum amplitude amplification and estimation},\ }\href {https://doi.org/10.1090/conm/305/05215} {\bibfield  {journal} {\bibinfo  {journal} {Quantum Computation and Information}\ ,\ \bibinfo {pages} {53–74}} (\bibinfo {year} {2002})}\BibitemShut {NoStop}%
\bibitem [{\citenamefont {Zhang}\ \emph {et~al.}(2023)\citenamefont {Zhang}, \citenamefont {Bosse},\ and\ \citenamefont {Cubitt}}]{zhang2023dissipative}%
  \BibitemOpen
  \bibfield  {author} {\bibinfo {author} {\bibfnamefont {D.}~\bibnamefont {Zhang}}, \bibinfo {author} {\bibfnamefont {J.~L.}\ \bibnamefont {Bosse}},\ and\ \bibinfo {author} {\bibfnamefont {T.}~\bibnamefont {Cubitt}},\ }\href@noop {} {\bibinfo {title} {Dissipative quantum gibbs sampling}} (\bibinfo {year} {2023}),\ \Eprint {https://arxiv.org/abs/2304.04526} {arXiv:2304.04526 [quant-ph]} \BibitemShut {NoStop}%
\bibitem [{\citenamefont {Rouzé}\ \emph {et~al.}(2024)\citenamefont {Rouzé}, \citenamefont {França},\ and\ \citenamefont {Álvaro M.~Alhambra}}]{rouze2024optimalquantumalgorithmgibbs}%
  \BibitemOpen
  \bibfield  {author} {\bibinfo {author} {\bibfnamefont {C.}~\bibnamefont {Rouzé}}, \bibinfo {author} {\bibfnamefont {D.~S.}\ \bibnamefont {França}},\ and\ \bibinfo {author} {\bibnamefont {Álvaro M.~Alhambra}},\ }\href {https://arxiv.org/abs/2411.04885} {\bibinfo {title} {Optimal quantum algorithm for gibbs state preparation}} (\bibinfo {year} {2024}),\ \Eprint {https://arxiv.org/abs/2411.04885} {arXiv:2411.04885 [quant-ph]} \BibitemShut {NoStop}%
\bibitem [{\citenamefont {Wu}\ and\ \citenamefont {Wang}(2022)}]{Wu_2022}%
  \BibitemOpen
  \bibfield  {author} {\bibinfo {author} {\bibfnamefont {Y.}~\bibnamefont {Wu}}\ and\ \bibinfo {author} {\bibfnamefont {J.~B.}\ \bibnamefont {Wang}},\ }\bibfield  {title} {\bibinfo {title} {Estimating gibbs partition function with quantum clifford sampling},\ }\href {https://doi.org/10.1088/2058-9565/ac47f0} {\bibfield  {journal} {\bibinfo  {journal} {Quantum Science and Technology}\ }\textbf {\bibinfo {volume} {7}},\ \bibinfo {pages} {025006} (\bibinfo {year} {2022})}\BibitemShut {NoStop}%
\bibitem [{\citenamefont {Matsumoto}\ \emph {et~al.}(2022)\citenamefont {Matsumoto}, \citenamefont {Shingu}, \citenamefont {Endo}, \citenamefont {Kawabata}, \citenamefont {Watabe}, \citenamefont {Nikuni}, \citenamefont {Hakoshima},\ and\ \citenamefont {Matsuzaki}}]{Matsumoto_2022}%
  \BibitemOpen
  \bibfield  {author} {\bibinfo {author} {\bibfnamefont {K.}~\bibnamefont {Matsumoto}}, \bibinfo {author} {\bibfnamefont {Y.}~\bibnamefont {Shingu}}, \bibinfo {author} {\bibfnamefont {S.}~\bibnamefont {Endo}}, \bibinfo {author} {\bibfnamefont {S.}~\bibnamefont {Kawabata}}, \bibinfo {author} {\bibfnamefont {S.}~\bibnamefont {Watabe}}, \bibinfo {author} {\bibfnamefont {T.}~\bibnamefont {Nikuni}}, \bibinfo {author} {\bibfnamefont {H.}~\bibnamefont {Hakoshima}},\ and\ \bibinfo {author} {\bibfnamefont {Y.}~\bibnamefont {Matsuzaki}},\ }\bibfield  {title} {\bibinfo {title} {Calculation of gibbs partition function with imaginary time evolution on near-term quantum computers},\ }\href {https://doi.org/10.35848/1347-4065/ac5152} {\bibfield  {journal} {\bibinfo  {journal} {Japanese Journal of Applied Physics}\ }\textbf {\bibinfo {volume} {61}},\ \bibinfo {pages} {042002} (\bibinfo {year} {2022})}\BibitemShut {NoStop}%
\bibitem [{\citenamefont {Lemieux}\ \emph {et~al.}(2020)\citenamefont {Lemieux}, \citenamefont {Heim}, \citenamefont {Poulin}, \citenamefont {Svore},\ and\ \citenamefont {Troyer}}]{Lemieux2020efficientquantum}%
  \BibitemOpen
  \bibfield  {author} {\bibinfo {author} {\bibfnamefont {J.}~\bibnamefont {Lemieux}}, \bibinfo {author} {\bibfnamefont {B.}~\bibnamefont {Heim}}, \bibinfo {author} {\bibfnamefont {D.}~\bibnamefont {Poulin}}, \bibinfo {author} {\bibfnamefont {K.}~\bibnamefont {Svore}},\ and\ \bibinfo {author} {\bibfnamefont {M.}~\bibnamefont {Troyer}},\ }\bibfield  {title} {\bibinfo {title} {Efficient {Q}uantum {W}alk {C}ircuits for {M}etropolis-{H}astings {A}lgorithm},\ }\href {https://doi.org/10.22331/q-2020-06-29-287} {\bibfield  {journal} {\bibinfo  {journal} {{Quantum}}\ }\textbf {\bibinfo {volume} {4}},\ \bibinfo {pages} {287} (\bibinfo {year} {2020})}\BibitemShut {NoStop}%
\bibitem [{\citenamefont {Wocjan}\ and\ \citenamefont {Temme}(2023)}]{Wocjan_2023}%
  \BibitemOpen
  \bibfield  {author} {\bibinfo {author} {\bibfnamefont {P.}~\bibnamefont {Wocjan}}\ and\ \bibinfo {author} {\bibfnamefont {K.}~\bibnamefont {Temme}},\ }\bibfield  {title} {\bibinfo {title} {Szegedy walk unitaries for quantum maps},\ }\href {https://doi.org/10.1007/s00220-023-04797-4} {\bibfield  {journal} {\bibinfo  {journal} {Communications in Mathematical Physics}\ }\textbf {\bibinfo {volume} {402}},\ \bibinfo {pages} {3201–3231} (\bibinfo {year} {2023})}\BibitemShut {NoStop}%
\bibitem [{\citenamefont {Low}\ and\ \citenamefont {Chuang}(2019)}]{LowChuangQuantum2019}%
  \BibitemOpen
  \bibfield  {author} {\bibinfo {author} {\bibfnamefont {G.~H.}\ \bibnamefont {Low}}\ and\ \bibinfo {author} {\bibfnamefont {I.~L.}\ \bibnamefont {Chuang}},\ }\bibfield  {title} {\bibinfo {title} {Hamiltonian {S}imulation by {Q}ubitization},\ }\href {https://doi.org/10.22331/q-2019-07-12-163} {\bibfield  {journal} {\bibinfo  {journal} {{Quantum}}\ }\textbf {\bibinfo {volume} {3}},\ \bibinfo {pages} {163} (\bibinfo {year} {2019})}\BibitemShut {NoStop}%
\bibitem [{\citenamefont {Gily\'{e}n}\ \emph {et~al.}(2019)\citenamefont {Gily\'{e}n}, \citenamefont {Su}, \citenamefont {Low},\ and\ \citenamefont {Wiebe}}]{Gilyen2019}%
  \BibitemOpen
  \bibfield  {author} {\bibinfo {author} {\bibfnamefont {A.}~\bibnamefont {Gily\'{e}n}}, \bibinfo {author} {\bibfnamefont {Y.}~\bibnamefont {Su}}, \bibinfo {author} {\bibfnamefont {G.~H.}\ \bibnamefont {Low}},\ and\ \bibinfo {author} {\bibfnamefont {N.}~\bibnamefont {Wiebe}},\ }\bibfield  {title} {\bibinfo {title} {Quantum singular value transformation and beyond: Exponential improvements for quantum matrix arithmetics},\ }in\ \href {https://doi.org/10.1145/3313276.3316366} {\emph {\bibinfo {booktitle} {Proceedings of the 51st Annual ACM SIGACT Symposium on Theory of Computing}}},\ \bibinfo {series and number} {STOC 2019}\ (\bibinfo  {publisher} {Association for Computing Machinery},\ \bibinfo {address} {New York, NY, USA},\ \bibinfo {year} {2019})\ p.\ \bibinfo {pages} {193–204}\BibitemShut {NoStop}%
\bibitem [{\citenamefont {de~Lima~Silva}\ \emph {et~al.}(2022)\citenamefont {de~Lima~Silva}, \citenamefont {Borges},\ and\ \citenamefont {Aolita}}]{silva2022fourierbased}%
  \BibitemOpen
  \bibfield  {author} {\bibinfo {author} {\bibfnamefont {T.}~\bibnamefont {de~Lima~Silva}}, \bibinfo {author} {\bibfnamefont {L.}~\bibnamefont {Borges}},\ and\ \bibinfo {author} {\bibfnamefont {L.}~\bibnamefont {Aolita}},\ }\href@noop {} {\bibinfo {title} {Fourier-based quantum signal processing}} (\bibinfo {year} {2022}),\ \Eprint {https://arxiv.org/abs/2206.02826} {arXiv:2206.02826 [quant-ph]} \BibitemShut {NoStop}%
\bibitem [{\citenamefont {Dong}\ \emph {et~al.}(2022)\citenamefont {Dong}, \citenamefont {Lin},\ and\ \citenamefont {Tong}}]{Dong_2022}%
  \BibitemOpen
  \bibfield  {author} {\bibinfo {author} {\bibfnamefont {Y.}~\bibnamefont {Dong}}, \bibinfo {author} {\bibfnamefont {L.}~\bibnamefont {Lin}},\ and\ \bibinfo {author} {\bibfnamefont {Y.}~\bibnamefont {Tong}},\ }\bibfield  {title} {\bibinfo {title} {Ground-state preparation and energy estimation on early fault-tolerant quantum computers via quantum eigenvalue transformation of unitary matrices},\ }\bibfield  {journal} {\bibinfo  {journal} {{PRX} Quantum}\ }\textbf {\bibinfo {volume} {3}},\ \href {https://doi.org/10.1103/prxquantum.3.040305} {10.1103/prxquantum.3.040305} (\bibinfo {year} {2022})\BibitemShut {NoStop}%
\bibitem [{\citenamefont {Melko}\ \emph {et~al.}(2029)\citenamefont {Melko}, \citenamefont {Carleo}, \citenamefont {Carrasquilla},\ and\ \citenamefont {Cirac}}]{Melko2019}%
  \BibitemOpen
  \bibfield  {author} {\bibinfo {author} {\bibfnamefont {R.~G.}\ \bibnamefont {Melko}}, \bibinfo {author} {\bibfnamefont {G.}~\bibnamefont {Carleo}}, \bibinfo {author} {\bibfnamefont {J.}~\bibnamefont {Carrasquilla}},\ and\ \bibinfo {author} {\bibfnamefont {J.~I.}\ \bibnamefont {Cirac}},\ }\bibfield  {title} {\bibinfo {title} {Restricted boltzmann machines in quantum physics},\ }\href {https://doi.org/10.1038/s41567-019-0545-1} {\bibfield  {journal} {\bibinfo  {journal} {Nature Physics}\ }\textbf {\bibinfo {volume} {15}},\ \bibinfo {pages} {887} (\bibinfo {year} {2029})}\BibitemShut {NoStop}%
\bibitem [{\citenamefont {Chowdhury}\ and\ \citenamefont {Somma}(2017)}]{Chowdhury2017}%
  \BibitemOpen
  \bibfield  {author} {\bibinfo {author} {\bibfnamefont {A.~N.}\ \bibnamefont {Chowdhury}}\ and\ \bibinfo {author} {\bibfnamefont {R.~D.}\ \bibnamefont {Somma}},\ }\bibfield  {title} {\bibinfo {title} {Quantum algorithms for gibbs sampling and hitting-time estimation},\ }\href@noop {} {\bibfield  {journal} {\bibinfo  {journal} {Quantum Info. Comput.}\ }\textbf {\bibinfo {volume} {17}},\ \bibinfo {pages} {41–64} (\bibinfo {year} {2017})}\BibitemShut {NoStop}%
\bibitem [{\citenamefont {Brandao}(2008)}]{brandao2008entanglement}%
  \BibitemOpen
  \bibfield  {author} {\bibinfo {author} {\bibfnamefont {F.~G. S.~L.}\ \bibnamefont {Brandao}},\ }\href@noop {} {\bibinfo {title} {Entanglement theory and the quantum simulation of many-body physics}} (\bibinfo {year} {2008}),\ \Eprint {https://arxiv.org/abs/0810.0026} {arXiv:0810.0026 [quant-ph]} \BibitemShut {NoStop}%
\bibitem [{\citenamefont {Brown}\ \emph {et~al.}(2001)\citenamefont {Brown}, \citenamefont {Cai},\ and\ \citenamefont {DasGupta}}]{Brown2001}%
  \BibitemOpen
  \bibfield  {author} {\bibinfo {author} {\bibfnamefont {L.~D.}\ \bibnamefont {Brown}}, \bibinfo {author} {\bibfnamefont {T.~T.}\ \bibnamefont {Cai}},\ and\ \bibinfo {author} {\bibfnamefont {A.}~\bibnamefont {DasGupta}},\ }\bibfield  {title} {\bibinfo {title} {{Interval Estimation for a Binomial Proportion}},\ }\href {https://doi.org/10.1214/ss/1009213286} {\bibfield  {journal} {\bibinfo  {journal} {Statistical Science}\ }\textbf {\bibinfo {volume} {16}},\ \bibinfo {pages} {101 } (\bibinfo {year} {2001})}\BibitemShut {NoStop}%
\bibitem [{\citenamefont {Pires}\ and\ \citenamefont {Amado}(2008)}]{pires2008interval}%
  \BibitemOpen
  \bibfield  {author} {\bibinfo {author} {\bibfnamefont {A.~M.}\ \bibnamefont {Pires}}\ and\ \bibinfo {author} {\bibfnamefont {C.}~\bibnamefont {Amado}},\ }\bibfield  {title} {\bibinfo {title} {Interval estimators for a binomial proportion: Comparison of twenty methods},\ }\href@noop {} {\bibfield  {journal} {\bibinfo  {journal} {REVSTAT-Statistical Journal}\ }\textbf {\bibinfo {volume} {6}},\ \bibinfo {pages} {165} (\bibinfo {year} {2008})}\BibitemShut {NoStop}%
\bibitem [{\citenamefont {Kikuchi}\ \emph {et~al.}(2023)\citenamefont {Kikuchi}, \citenamefont {Mc~Keever}, \citenamefont {Coopmans}, \citenamefont {Lubasch},\ and\ \citenamefont {Benedetti}}]{Kikuchi_2023}%
  \BibitemOpen
  \bibfield  {author} {\bibinfo {author} {\bibfnamefont {Y.}~\bibnamefont {Kikuchi}}, \bibinfo {author} {\bibfnamefont {C.}~\bibnamefont {Mc~Keever}}, \bibinfo {author} {\bibfnamefont {L.}~\bibnamefont {Coopmans}}, \bibinfo {author} {\bibfnamefont {M.}~\bibnamefont {Lubasch}},\ and\ \bibinfo {author} {\bibfnamefont {M.}~\bibnamefont {Benedetti}},\ }\bibfield  {title} {\bibinfo {title} {Realization of quantum signal processing on a noisy quantum computer},\ }\bibfield  {journal} {\bibinfo  {journal} {npj Quantum Information}\ }\textbf {\bibinfo {volume} {9}},\ \href {https://doi.org/10.1038/s41534-023-00762-0} {10.1038/s41534-023-00762-0} (\bibinfo {year} {2023})\BibitemShut {NoStop}%
\bibitem [{\citenamefont {Temme}\ \emph {et~al.}(2017)\citenamefont {Temme}, \citenamefont {Bravyi},\ and\ \citenamefont {Gambetta}}]{temme2017zne}%
  \BibitemOpen
  \bibfield  {author} {\bibinfo {author} {\bibfnamefont {K.}~\bibnamefont {Temme}}, \bibinfo {author} {\bibfnamefont {S.}~\bibnamefont {Bravyi}},\ and\ \bibinfo {author} {\bibfnamefont {J.~M.}\ \bibnamefont {Gambetta}},\ }\bibfield  {title} {\bibinfo {title} {Error mitigation for short-depth quantum circuits},\ }\href {https://doi.org/10.1103/PhysRevLett.119.180509} {\bibfield  {journal} {\bibinfo  {journal} {Phys. Rev. Lett.}\ }\textbf {\bibinfo {volume} {119}},\ \bibinfo {pages} {180509} (\bibinfo {year} {2017})}\BibitemShut {NoStop}%
\bibitem [{\citenamefont {Li}\ and\ \citenamefont {Benjamin}(2017)}]{li2017zne}%
  \BibitemOpen
  \bibfield  {author} {\bibinfo {author} {\bibfnamefont {Y.}~\bibnamefont {Li}}\ and\ \bibinfo {author} {\bibfnamefont {S.~C.}\ \bibnamefont {Benjamin}},\ }\bibfield  {title} {\bibinfo {title} {Efficient variational quantum simulator incorporating active error minimization},\ }\href {https://doi.org/10.1103/PhysRevX.7.021050} {\bibfield  {journal} {\bibinfo  {journal} {Phys. Rev. X}\ }\textbf {\bibinfo {volume} {7}},\ \bibinfo {pages} {021050} (\bibinfo {year} {2017})}\BibitemShut {NoStop}%
\bibitem [{\citenamefont {Vugrin}\ \emph {et~al.}(2007)\citenamefont {Vugrin}, \citenamefont {Swiler}, \citenamefont {Roberts}, \citenamefont {Stucky-Mack},\ and\ \citenamefont {Sullivan}}]{Vugrin2007}%
  \BibitemOpen
  \bibfield  {author} {\bibinfo {author} {\bibfnamefont {K.~W.}\ \bibnamefont {Vugrin}}, \bibinfo {author} {\bibfnamefont {L.~P.}\ \bibnamefont {Swiler}}, \bibinfo {author} {\bibfnamefont {R.~M.}\ \bibnamefont {Roberts}}, \bibinfo {author} {\bibfnamefont {N.~J.}\ \bibnamefont {Stucky-Mack}},\ and\ \bibinfo {author} {\bibfnamefont {S.~P.}\ \bibnamefont {Sullivan}},\ }\bibfield  {title} {\bibinfo {title} {Confidence region estimation techniques for nonlinear regression in groundwater flow: Three case studies},\ }\href {https://doi.org/https://doi.org/10.1029/2005WR004804} {\bibfield  {journal} {\bibinfo  {journal} {Water Resources Research}\ }\textbf {\bibinfo {volume} {43}},\ \bibinfo {pages} {W03423} (\bibinfo {year} {2007})}\BibitemShut {NoStop}%
\bibitem [{\citenamefont {Ku}(1966)}]{Ku1966}%
  \BibitemOpen
  \bibfield  {author} {\bibinfo {author} {\bibfnamefont {H.}~\bibnamefont {Ku}},\ }\bibfield  {title} {\bibinfo {title} {Notes on the use of propagation of error formulas},\ }\href {https://doi.org/https://dx.doi.org/10.6028/jres.070C.025} {\bibfield  {journal} {\bibinfo  {journal} {Journal of Research of the National Bureau of Standards, Section C: Engineering and Instrumentation}\ }\textbf {\bibinfo {volume} {70C}},\ \bibinfo {pages} {263} (\bibinfo {year} {1966})}\BibitemShut {NoStop}%
\bibitem [{\citenamefont {Vovrosh}\ \emph {et~al.}(2021)\citenamefont {Vovrosh}, \citenamefont {Khosla}, \citenamefont {Greenaway}, \citenamefont {Self}, \citenamefont {Kim},\ and\ \citenamefont {Knolle}}]{vovorosh2021_mitigation}%
  \BibitemOpen
  \bibfield  {author} {\bibinfo {author} {\bibfnamefont {J.}~\bibnamefont {Vovrosh}}, \bibinfo {author} {\bibfnamefont {K.~E.}\ \bibnamefont {Khosla}}, \bibinfo {author} {\bibfnamefont {S.}~\bibnamefont {Greenaway}}, \bibinfo {author} {\bibfnamefont {C.}~\bibnamefont {Self}}, \bibinfo {author} {\bibfnamefont {M.~S.}\ \bibnamefont {Kim}},\ and\ \bibinfo {author} {\bibfnamefont {J.}~\bibnamefont {Knolle}},\ }\bibfield  {title} {\bibinfo {title} {Simple mitigation of global depolarizing errors in quantum simulations},\ }\href {https://doi.org/10.1103/PhysRevE.104.035309} {\bibfield  {journal} {\bibinfo  {journal} {Phys. Rev. E}\ }\textbf {\bibinfo {volume} {104}},\ \bibinfo {pages} {035309} (\bibinfo {year} {2021})}\BibitemShut {NoStop}%
\bibitem [{\citenamefont {Sachdeva}\ and\ \citenamefont {Vishnoi}(2013)}]{sachdeva_approximation_2013}%
  \BibitemOpen
  \bibfield  {author} {\bibinfo {author} {\bibfnamefont {S.}~\bibnamefont {Sachdeva}}\ and\ \bibinfo {author} {\bibfnamefont {N.}~\bibnamefont {Vishnoi}},\ }\href {https://doi.org/10.48550/arXiv.1309.4882} {\bibinfo {title} {Approximation {Theory} and the {Design} of {Fast} {Algorithms}}} (\bibinfo {year} {2013}),\ \bibinfo {note} {arXiv:1309.4882 [cs, math]}\BibitemShut {NoStop}%
\bibitem [{\citenamefont {Somma}\ \emph {et~al.}(2008)\citenamefont {Somma}, \citenamefont {Boixo}, \citenamefont {Barnum},\ and\ \citenamefont {Knill}}]{Somma_2008}%
  \BibitemOpen
  \bibfield  {author} {\bibinfo {author} {\bibfnamefont {R.~D.}\ \bibnamefont {Somma}}, \bibinfo {author} {\bibfnamefont {S.}~\bibnamefont {Boixo}}, \bibinfo {author} {\bibfnamefont {H.}~\bibnamefont {Barnum}},\ and\ \bibinfo {author} {\bibfnamefont {E.}~\bibnamefont {Knill}},\ }\bibfield  {title} {\bibinfo {title} {Quantum simulations of classical annealing processes},\ }\bibfield  {journal} {\bibinfo  {journal} {Physical Review Letters}\ }\textbf {\bibinfo {volume} {101}},\ \href {https://doi.org/10.1103/physrevlett.101.130504} {10.1103/physrevlett.101.130504} (\bibinfo {year} {2008})\BibitemShut {NoStop}%
\bibitem [{\citenamefont {Silva}\ \emph {et~al.}(2023)\citenamefont {Silva}, \citenamefont {Taddei}, \citenamefont {Carrazza},\ and\ \citenamefont {Aolita}}]{silva_fragmented_2022}%
  \BibitemOpen
  \bibfield  {author} {\bibinfo {author} {\bibfnamefont {T.~L.}\ \bibnamefont {Silva}}, \bibinfo {author} {\bibfnamefont {M.~M.}\ \bibnamefont {Taddei}}, \bibinfo {author} {\bibfnamefont {S.}~\bibnamefont {Carrazza}},\ and\ \bibinfo {author} {\bibfnamefont {L.}~\bibnamefont {Aolita}},\ }\bibfield  {title} {\bibinfo {title} {Fragmented imaginary-time evolution for early-stage quantum signal processors},\ }\href {https://doi.org/10.1038/s41598-023-45540-2} {\bibfield  {journal} {\bibinfo  {journal} {Scientific Reports}\ }\textbf {\bibinfo {volume} {13}},\ \bibinfo {pages} {18258} (\bibinfo {year} {2023})}\BibitemShut {NoStop}%
\bibitem [{Ion()}]{IonQ}%
  \BibitemOpen
  \href@noop {} {\bibinfo {title} {{Getting started with Native Gates}}},\ \bibinfo {howpublished} {\url{https://ionq.com/docs/getting-started-with-native-gates}},\ \bibinfo {note} {accessed: 2023-08-20}\BibitemShut {NoStop}%
\bibitem [{\citenamefont {Wang}\ \emph {et~al.}(2023)\citenamefont {Wang}, \citenamefont {França}, \citenamefont {Rendon},\ and\ \citenamefont {Johnson}}]{wang2023fastergroundstateenergy}%
  \BibitemOpen
  \bibfield  {author} {\bibinfo {author} {\bibfnamefont {G.}~\bibnamefont {Wang}}, \bibinfo {author} {\bibfnamefont {D.~S.}\ \bibnamefont {França}}, \bibinfo {author} {\bibfnamefont {G.}~\bibnamefont {Rendon}},\ and\ \bibinfo {author} {\bibfnamefont {P.~D.}\ \bibnamefont {Johnson}},\ }\href {https://arxiv.org/abs/2304.09827} {\bibinfo {title} {Faster ground state energy estimation on early fault-tolerant quantum computers via rejection sampling}} (\bibinfo {year} {2023}),\ \Eprint {https://arxiv.org/abs/2304.09827} {arXiv:2304.09827 [quant-ph]} \BibitemShut {NoStop}%
\end{thebibliography}%

\appendix

\section{Proofs of theorems and lemmas}

\subsection{Proof of Theorem \ref{thm:prob}}\label{sec:proof_prob}

Suppose that of the total number of coin flips, $\mathcal{S}_{\suc}$ of them were successes. Agresti-Coull interval \cite{Brown2001} can be used to obtain an estimate for the probability of the coin as 
\begin{equation}\label{eq:p_suc_estimator}
    \hat{p}_{\suc}^{(\beta)}=\frac{1}{\mathcal{S}+z_{\delta}^2}\left(\mathcal{S}_{\suc}+\frac{z_{\delta}^2}{2}\right) 
\end{equation}
up to an additive error
\begin{equation}\label{eq:error_p}
    \varepsilon_p \leq z_{\delta}\sqrt{\frac{\hat{p}_{\suc}^{(\beta)}}{\mathcal{S}}(1-\hat{p}_{\suc}^{(\beta)})}
\end{equation}
 with confidence $1-\delta$, where $z_{\delta}$ is the quantile of a standard normal distribution at $1-\delta/2$, i.e., it satisfies $\frac{1}{\sqrt{2\pi}}\int_{-\infty}^{z_\delta}e^{-t^2/2}dt=\frac{\delta}{2}$. $z_{\delta}$ is a slow-varying function of the confidence. For instance, $z_{.05}=1.96$, while $z_{10^{-9}}=6.11$. 
 
 From the success probability estimate, an estimate for the partition function is obtained using Eq. \eqref{eq:part_prob}. The choice of precision $\varepsilon_p\leq\frac{Z_\beta}{2^n e^\beta}\,\frac{\varepsilon_r}{2}$ ensures a target relative precision $\varepsilon_r/2$ for the partition function estimate. The required number of samples is then given by Eq. \eqref{eq:error_p} as 
\begin{equation}
    \mathcal{S} \geq 8\,\frac{z_{\delta}^2}{\varepsilon_r^2}\,\frac{2^ne^\beta}{Z_\beta},
\end{equation}
where we also used Eq.\ \eqref{eq:part_prob} and $(1+\varepsilon_r)<2$. 

For simplicity, so far, we have considered that $V_{f_\beta}$ implements $e^{-\frac{\beta}{2} H}$  exactly. The most we can hope for is that $V_{f_\beta}$ satisfies Eq.\ \eqref{eq:qsp}  and, hence, is an $\varepsilon'$-approximation of the exponential function. 
 We denote $\Tilde{p}_{\suc}^{(\beta)}$ as the actual probability of success, given the approximate implementation, and keep $p_{\suc}^{(\beta)}$ for the ideal implementation. The difference between the two satisfy:
\begin{align}\label{eq:ptilde}
 \nonumber   |\Tilde{p}_{\suc}^{(\beta)}-p_{\suc}^{(\beta)}|&=\frac{e^{-\beta }}{2^n}\left|\Tr\left[\Tilde{f}_\beta[H]^2-e^{-\beta H}\right]\right|\\ \nonumber
     & \leq\frac{1}{2^n}\sum_\lambda\left|\left(e^{-\frac{\beta}{2}}\Tilde{f}_\beta(\lambda)\right)^2-e^{-\beta}e^{-\beta \lambda}\right|\\ \nonumber
     & \leq\frac{1}{2^n}\sum_\lambda\left|\left(e^{-\frac{\beta}{2}}e^{-\frac{\beta}{2} \lambda}+\varepsilon'\right)^2-e^{-\beta}e^{-\beta \lambda}\right|\\ \nonumber
     & =\frac{1}{2^n}\sum_\lambda\left|2\,e^{-\frac{\beta}{2}}e^{-\frac{\beta}{2} \lambda}\varepsilon'+\varepsilon'^2\right|\\
     & \leq \frac{1}{2^n} 3\times 2^n\varepsilon'=3\varepsilon'.
\end{align}
The sum in $\lambda$ runs over the eigenvalues of $H$, and we used Eq.\ \eqref{eq:qsp} in the third line and $\varepsilon'< 1$ in the fifth line. The error in the probability of success leads to a bias $3\,e^\beta 2^n\varepsilon'$ in the statistical estimator of the partition function. If we choose $\varepsilon'\leq\frac{Z_\beta}{6\,e^\beta2^n}\varepsilon_r$, then the relative error due to the approximation is kept as $\varepsilon_r/2$, and the total error (approximation bias + statistical error) is at most $\varepsilon_r$.

\subsection{Proof of Lemma \ref{lem:part_succ}}

The probability that a success will occur only in the $R$-th run of the circuit is given as $\text{Pr}(R)=(1-p_{\suc}^{(\beta)})^{R-1}p_{\suc}^{(\beta)}$, which is the probability of getting $R-1$ consecutive failures times the probability of obtaining one success. Therefore, the average number of repetitions to get one success is
\begin{equation}\label{eq:mean_R}
    \Bar{R}=\sum_{r=1}^\infty p_{\suc}^{(\beta)}\,(1-p_{\suc}^{(\beta)})^{r-1} \,r =\frac{1}{p_{\suc}^{(\beta)}},
\end{equation}
by using that $(1-x)^{-1}=\sum_{r=0}^\infty x^r$. The result follows by using Eq.\ \eqref{eq:part_prob}.

\subsection{Proof of Theorem \ref{thm:trials}}

\begin{proof}
One can show that
\begin{equation}
    \overline{R^2}=\sum_{r=1}^\infty p_{\suc}^{(\beta)}\,(1-p_{\suc}^{(\beta)})^{r-1} \,r^2 =\frac{2-p_{\suc}^{(\beta)}}{\big(p_{\suc}^{(\beta)}\big)^2}.
\end{equation}
Combining the above with Eq. \eqref{eq:mean_R} gives the variance
\begin{equation}
    \text{Var}(R)=\overline{R^2}-\Bar{R}^2=\frac{1-p_{\suc}^{(\beta)}}{\big(p_{\suc}^{(\beta)}\big)^2}.
\end{equation}

Chebyshev inequality then yields the number of samples of the random variable to attain an additive precision $\varepsilon$ with confidence $(1-\delta)$ as
\begin{equation}\label{eq:sample}
    S_{\suc}=\frac{1-p_{\suc}^{(\beta)}}{\delta\, \varepsilon^2\,\big(p_{\suc}^{(\beta)}\big)^2}.
\end{equation}

From the empirical average $\Hat{\Bar{R}}=\frac{1}{S_{\suc}}\sum_{j=1}^{S_{\suc}} R_j$, using Eqs. \eqref{eq:mean_R} and \eqref{eq:part_prob}, we obtain an estimate for the partition function  as $\Hat{Z}_\beta=\frac{2^n\,e^\beta}{\Hat{\Bar{R}}}$. An additive error $\varepsilon$ in $\Hat{\Bar{R}}$ propagates to the partition function estimate as an additive error
\begin{equation}\label{eq:errorR}
    \varepsilon_a=\left|\frac{d\Hat{Z}_\beta}{d\Hat{\Bar{R}}}\Bigg|_{\Bar{R}}\varepsilon\right|=\frac{2^n e^\beta}{\Bar{R}^2}\varepsilon=Z_\beta \,p_{\suc}\,\varepsilon,
\end{equation}
which is easily transformed into a relative precision by making $\varepsilon=\varepsilon_r/p_{\suc}^{(\beta)}$. In turn, combined with Eq.\ \eqref{eq:sample} it leads to 
\begin{equation}
    S_{\suc}=\frac{1-p_{\suc}^{(\beta)}}{\delta\, \varepsilon_r^2}\leq \frac{1}{\delta\, \varepsilon_r^2}.
\end{equation}
Finally, from the average circuit runs to obtain a success  $\Bar{R}=\frac{1}{p_{\suc}^{(\beta)}}=\frac{2^n\,e^\beta}{Z_\beta}$, we obtain the average total number of circuit uses as $S_{\suc}\Bar{R}=(2^n e^\beta)/\left(\delta\,\varepsilon_r^2 Z_\beta\right)$.  
\end{proof}

\end{document}